\documentclass{emulateapj}

\usepackage{natbib}
\usepackage{amssymb,amsmath}
\usepackage{color,hyperref}
\definecolor{linkcolor}{rgb}{0,0,0.5}
\hypersetup{colorlinks=true,linkcolor=linkcolor,citecolor=linkcolor,
            filecolor=linkcolor,urlcolor=linkcolor}
\usepackage{url}

\newcommand{\figref}[1]{\ref{fig:#1}}
\newcommand{\Fig}[1]{\figurename~\figref{#1}}
\newcommand{\fig}[1]{\Fig{#1}}
\newcommand{\figlabel}[1]{\label{fig:#1}}
\newcommand{\Tab}[1]{Table~\ref{tab:#1}}
\newcommand{\tab}[1]{\Tab{#1}}
\newcommand{\tablabel}[1]{\label{tab:#1}}
\newcommand{\Eq}[1]{Equation~(\ref{eq:#1})}
\newcommand{\eq}[1]{\Eq{#1}}

\newcommand{\eqlabel}[1]{\label{eq:#1}}
\newcommand{\sectionname}{Section}
\newcommand{\Sect}[1]{\sectionname~\ref{sect:#1}}
\newcommand{\sect}[1]{\Sect{#1}}

\newcommand{\sectlabel}[1]{\label{sect:#1}}

\newcommand{\ntotal}{7470}
\newcommand{\nfail}{414}
\newcommand{\ncalc}{7056}

\newcommand{\nbadphot}{28} 
\newcommand{\nbadrowemcmc}{233}
\newcommand{\nbadstellar}{74} 
\newcommand{\nbadsec}{8} 
\newcommand{\nbadtrapfit}{38} 
\newcommand{\nbadroche}{39}

\newcommand{\nlongfp}{526}

\newcommand{\nreliable}{2857} 
\newcommand{\nval}{1935} 
\newcommand{\nreliableFP}{419} 
\newcommand{\nvalnew}{1284} 
\newcommand{\nfpnew}{428}  
\newcommand{\nhz}{9}

\newcommand{\posprobthresh}{0.3}

\newcommand{\kepler}{\textit{Kepler}}
\newcommand{\vespa}{\texttt{vespa}}
\newcommand{\isochrones}{\texttt{isochrones}}
\newcommand{\bvec}[1]{{\ensuremath{\boldsymbol{#1}}}}

\newcommand{\teff}{\ensuremath{T_{\rm eff}}}
\newcommand{\logg}{\ensuremath{\log{g}}}

\newcommand{\rsun}{\ensuremath{R_\sun}}
\newcommand{\msun}{\ensuremath{M_\sun}}

\newcommand{\rstar}{\ensuremath{R_\star}}
\newcommand{\mstar}{\ensuremath{M_\star}}

\newcommand{\rpl}{\ensuremath{R_{\rm P}}}

\newcommand{\rearth}{\ensuremath{R_{\earth}}}

\defcitealias{Morton:2012}{M12}
\defcitealias{Huber:2014}{H14}
\defcitealias{Dressing:2013}{D13}

\shorttitle{False Positive Probabilities for all KOIs}
\shortauthors{Morton et al.}

\begin{document}


\title{False Positive Probabilities for all Kepler Objects of Interest: \\
        \nvalnew\ newly validated planets and \nfpnew\ likely false positives}


\author{Timothy D. Morton\altaffilmark{1}, Stephen T. Bryson\altaffilmark{2}, Jeffrey L. Coughlin\altaffilmark{2,3}, Jason F. Rowe\altaffilmark{4}, Ganesh Ravichandran\altaffilmark{5}, Erik A. Petigura\altaffilmark{6,7}, 
Michael R. Haas\altaffilmark{2}, and Natalie M. Batalha\altaffilmark{2}}


\altaffiltext{1}{Department of Astrophysical Sciences, Princeton University, 4 Ivy Lane, Princeton, NJ 08544, USA, tdm@astro.princeton.edu}
\altaffiltext{2}{NASA Ames Research Center, M/S 244-30, Moffett Field, CA 94035, USA}
\altaffiltext{3}{SETI Institute, 189 Bernardo Ave, Mountain View, CA 94043, USA}
\altaffiltext{4}{D\'epartement de Physique, Universit\'e de Montr\'eal, Montr\'eal, QC, Canada, H3T 1J4}
\altaffiltext{5}{Department of Computer Science, Columbia University, 1214 Amsterdam Ave, New York, NY 10027}
\altaffiltext{6}{California Institute of Technology, Pasadena, CA 91125, USA}
\altaffiltext{7}{Hubble Fellow}



\begin{abstract}
We present astrophysical false positive probability calculations for
every \kepler\ Object of Interest (KOI)---the first large-scale
demonstration of a fully automated transiting planet validation
procedure.  Out of \ncalc\ KOIs, we determine that \nval\ have
probabilities $<$1\% to be astrophysical false positives, and thus may
be considered validated planets.  \nvalnew\ of these have not yet been
validated or confirmed by other methods.  In addition, we identify
\nfpnew\ KOIs likely to be false positives that have not yet been
identified as such, though some of these may be a result of
unidentified transit timing variations. A side product of these
calculations is full stellar property posterior samplings for every
host star, modeled as single, binary, and triple systems.  These
calculations use \vespa, a publicly available Python package able to
be easily applied to any transiting exoplanet candidate.
\end{abstract}




\section{Introduction}

The \kepler\ mission has revolutionized our understanding of
exoplanets.  Among many other important discoveries, \kepler\ has
identified several previously unsuspected features of planetary
systems, such as the prevalence of planets between the size of Earth
and Neptune, and a population of very compact multiple-planet systems.
And perhaps most notably, it has enabled for the first time estimates
of the occurrence rates of small planets ($\gtrsim$1 $R_\oplus$) out
to orbits of about one year
\citep[e.g.][]{Petigura:2013,DFM:2014,Burke:2015}.  It is important to
remember, however, that these revolutionary discoveries depend
intimately on another revolution---how to interpret transiting planet
\textit{candidate} signals in the absence of unambiguous positive
confirmation of their veracity.

Before \kepler, every survey searching for transiting exoplanets
demanded that a candidate signal be verified as a true planet via
radial velocity (RV) measurement of its mass.  This would involve a
series of follow-up observations in order to weed out astrophysical
false positive scenarios---typically stellar eclipsing binaries in
various configurations.  However, following this model has been
largely impossible for \kepler\ because of the quantity and character
of the planet candidates (thousands of mostly small-planet candidates
around relatively faint stars).  There have been a small number of
\kepler\ planets with masses measured by RVs
\citep[e.g.,][]{Marcy:2014,Santerne:2015}, and significantly more that
have been confirmed as planets by measurement of transit timing
variations (TTVs) in multi-planet systems
\citep[e.g.,][]{Ford:2012,Steffen:2012,Fabrycky:2012,Steffen:2013,Jontof:2015},
but this still leaves the vast majority of candidates inaccessible to
dynamical confirmation.

This situation has inspired the development of \emph{probabilistic
validation} as a new approach to evaluating transit candidates.  The
principle of probabilistic validation is to demonstrate that all
conceivable astrophysical false positive scenarios are negligibly
likely to be the cause of a transit candidate signal compared to the
explanation of a planet transiting the presumed target star.  The
\verb|BLENDER| method pioneered this approach and has validated many
\kepler\ candidates
\citep[e.g.,][]{Borucki:2012,Kipping:2014,Torres:2015}.   More
recently, the \verb|PASTIS| analysis suite has been introduced
\citep{Diaz:2014} and used to validate both \kepler\ and
\textit{CoRoT} candidates \citep[e.g.,][]{Santerne:2014,Moutou:2014}.
An alternative validation approach for candidates in multiple-planet
systems has also been applied to a large number of \kepler\ systems
based on the general argument that it is unlikely to see multiple
false-positive signals in the same \kepler\ light curve
\citep{Lissauer:2012}, resulting resulting in validations of over 800
planets with 99\% confidence \citep{Lissauer:2014,Rowe:2014}. This
methodology differs from the \verb|BLENDER|/\verb|PASTIS| approach in
two significant ways: (a) it is applicable only to planets in multi-
planet systems, and (b) it relies on broad-brush general arguments
rather than analyzing the details of candidate signals individually.

While they have both proven useful for the purposes of validating
individual candidates of particular interest, neither \verb|BLENDER|
nor \verb|PASTIS| is designed for fully automated batch processing of
large numbers of candidates.  \citet{Morton:2012} describes a
computationally simpler planet validation procedure designed for
exactly such a purpose, based on the idea of describing eclipse light
curves as simple trapezoids and simulating realistic populations of
astrophysical false positives. This procedure has also been used in
the literature to validate a number of \kepler\ planets
\citep[e.g.,][]{Muirhead:2012,Dawson:2012,Swift:2013}, and has also
been applied to a number of candidates found by the \textit{K2}
mission \citep{Montet:2015,Becker:2015}.  The code that implements
this procedure is publicly available in the Python module
\vespa\footnote{\url{https://github.com/timothydmorton/vespa}}
\citep{vespa}.

This work presents results of applying \vespa\ \emph{en masse} to the
entire \kepler\ catalog. This is both the first time that most
\kepler\ candidates have been individually analyzed to assess false
positive probability and the first time that a detailed automated
planet validation calculation has been applied on such a large scale.
\Sect{methods} describes the methods used, \sect{data} describes the data set,
\sect{results} presents the results, \sect{comparisons} compares these
results with observational studies, and \sect{conclusions} contains concluding
remarks.


\section{Methods}
\sectlabel{methods}

In this work, we apply the fully automated FPP-computing procedure
described in \citet[][hereafter \citetalias{Morton:2012}]{Morton:2012}
to \ncalc\ Kepler Objects of Interest (KOIs; see \sect{data} for
details).  While we refer the reader to \citetalias{Morton:2012} for a
detailed description of the method, we outline it briefly in this
section.  


\subsection{False Positive Probabilities}
\sectlabel{methods:fpp}

The basic idea of \vespa\ is to assign probabilities to different
hypotheses that might describe a transiting planet
candidate signal.  If $\{H_i\}$ is the set of all considered
hypotheses, the probability for any given model $i$ is
\begin{equation}
  \eqlabel{prob}
  \mathrm{Pr}\left(H_i\right) = \frac{\pi_i \mathcal
    L_i}{\displaystyle \sum_j \pi_j \mathcal L_j},
\end{equation}
where $\pi_i$ is the ``hypothesis prior'' and $\mathcal L_i$ is the
``hypothesis likelihood''\footnote{This factor is more widely known as
  the ``Bayesian evidence'' or ``marginalized likelihood'';
  \citet{Morton:2014b} argues for the term ``hypothesis likelihood,''
  as it can be clarifying to think of it that way.}
The prior represents how intrinsically probable the hypothesized
scenario is to exist, and the likelihood represents how closely the
shape of the observed transit signal matches with the expected shape
of a signal produced by the hypothesis.

The \vespa\ procedure models an eclipse signal as a simple trapezoid,
parametrized by depth $\delta$, total duration $T$, and shape
parameter $T / \tau$, where $\tau$ is the ``ingress/egress'' duration
(such that a completely V-shaped transit has $T/\tau = 2$).  For the
transit signal being evaluated, the joint posterior probability
density function (PDF) of these shape parameters is sampled with
Markov Chain Monte Carlo (MCMC), using the \texttt{emcee} sampler
\citep{emcee}.  This allows the likelihood for each hypothesis to be
determined by simulating a physically realistic population of the
hypothesized astrophysical scenario and using this population to
define the PDF for the trapezoidal parameters under the hypothesis.
The likelihood is then \begin{equation}   \eqlabel{lhood}   \mathcal
L_i = \displaystyle \int p_\mathrm{sig}\left(\bvec{\theta}\right)
p_i\left(\bvec{\theta}\right)\,d\bvec{\theta}, \end{equation} where
$\bvec{\theta}$ is the vector of trapezoidal shape parameters,
$p_\mathrm{sig}$ is the posterior PDF of the signal, and $p_i$ is the
PDF for the parameters under hypothesis $i$.\footnote{$\mathcal L_i$
may be seen to be the ``evidence'' or ``marginalized likelihood'' of
the trapezoidal model under hypothesis $i$, with $p_\mathrm{sig}$
being the likelihood and $p_i$ being the prior, integrated over the
$\bvec{\theta}$ parameter space.  But for clarity, and for continuity
with previous publications, we continue to call $\mathcal L_i$ the
``likelihood'' for hypothesis $i$.}  The hypotheses supported by
\vespa\ are the following: unblended eclipsing binary (EB),
hierarchical-triple eclipsing binary (HEB), chance-aligned
background/foreground eclipsing binary (BEB), and transiting planet
(Pl).\footnote{We note that we do not consider ``blended transiting planet''
false positive scenarios, either due to physically associated or 
chance-aligned companions.  See \sect{btp} for more discussion.} 
In this work, we also implement ``double-period'' versions of
each of the stellar false positive scenarios, acknowledging the
possibility that if an eclipsing binary has similar primary and
secondary eclipse depths, then it might be mischaracterized as a
primary-only transiting planet signal at twice the orbital period
(especially if diluted).  We note that the determination of the
diluted eclipse depth of all these blended scenarios assumes that the
light from the blended system is \emph{fully contained} within the
target's photometric aperture.  That is, these scenarios do not
account for the possibility that only a small fraction of the light
from a nearby contaminating star might be in the aperture, many of
which have already been identified via other methods
\cite{Bryson:2013,Coughlin:2014}.

Observational constraints are incorporated in two different ways.
First, photometric (or spectroscopic/asteroseismic) measurements of
the target star are folded into the population simulations of each
hypothesis (see \sect{methods:stellar}).  All other constraints are
applied to narrow down which simulated instances of each scenario may
be counted in the final prior and likelihood evalulations; for
example, only blended eclipsing binaries with secondary eclipse depths
shallower than the observed limits contribute to the construction of
the $p_i$ trapezoidal shape parameter PDF.  For the ``double-period''
scenarios, we require the primary and secondary eclipse signals to
have depths within 3-$\sigma$ of each other, where $\sigma$ is defined
as the fitted uncertainty in the trapezoid-model depth of the
candidate signal.

The steps \vespa\ takes to calculate the FPP of a transit signal
are thus as follows:
\begin{enumerate}
\item Generate posterior samples for the transit signal under the
  trapezoid model, using MCMC.
\item Generate population simulations for each hypothesis scenario
  being considered (conditioned on available observations of the
  target star; see \sect{methods:stellar}).
\item Fit each simulated eclipse in each scenario with a trapezoid
  model (using least-squares optimization).
\item Evaluate priors and likelihoods for each hypothesis, taking into
  account all available observational constraints.
\item Use \eq{prob} to calculate the posterior probability for each scenario.
\end{enumerate}

To quantify uncertainty due to the Monte Carlo nature of this procedure, 
\vespa\ is also able to repeat these calculations any desired number of times 
by bootstrap resampling (with replacement) of the simulated populations, 
and recalculating the likelihoods based on the resampled populations.  
This mitigates the chances for rare outliers in a simulation to significantly 
affect the calculated FPP.

We note that a built-in weakness of model selection is that it assumes that
the set of models being considered is comprehensive.  This could
in principle lead to a situation where one model is strongly preferred
over all other models, but even that model is a poor explanation of
the data---in this work, this could lead to improperly validated planets.  
There are two general strategies to try to address this issue.  The first 
is to somehow quantify the absolute goodness-of-fit of the models, and 
require that a validated planet pass some threshold test.  The other 
strategy, that we adopt here, is to expand the set of models to be more 
comprehensive.  In order to do this, we introduce two artificial models:
``boxy'' and ``long.''  The ``boxy'' likelihood function is a step function
at some minimum value of $T/\tau$ (zero below this threshold, and constant
above), and constant throughout the space of the other trapezoidal parameters.
Similarly, the likelihood of the ``long'' model is a step function at some
minimum threshold value of duration $T$.  These thresholds  
are set relative to the simulated trapezoidal shapes of the planet model:
the $T/\tau$ threshold is the maximum value from the simulated planet population,
and the $T$ threshold is the 99\% percentile of simulated planet
population values.  We also choose the model priors for these artificial
models to be low, reflecting that we expect only a small number of signals
to be unexplained by any of the astrophysical scenarios: the number that we
choose for each of these models is $5 \times 10^{-5}$, corresponding to an expectation
that there may be $\sim$10 such signals among the $\sim$200,000 \kepler\ targets.


\subsection{Stellar Properties}
\sectlabel{methods:stellar}

The most substantial difference between the current implementation of
\vespa\ and the procedure documented in \citetalias{Morton:2012} is
how stellar properties are treated.  Previously, either the target
star's mass and radius were explicitly provided, or they were randomly
generated according to the stellar population expected along the line
of sight by the TRIdimensional modeL of thE GALaxy (TRILEGAL) Galactic
stellar population synthesis tool, but constrained to agree with some
observed color(s) of the star (e.g. $J-K$), to within some specified
tolerance.  This strategy was used both to generate the host stars for
the transiting planet model and the binary and triple stars for the EB
and HEB false positive models.

The new method now used by \vespa\ uses the \isochrones\ Python module
\citep{isochrones} to fold in observational constraints on the host
star.  At its core, \isochrones\ performs 3-D linear interpolation in
mass--[Fe/H]--age parameter space for a given stellar model grid.
This method of stellar modeling for FPP calculation debuted in
\citet{Montet:2015} and is explained there in more detail.  Instead of
randomly generating stars (or binary or triple systems of stars) from
a predefined distribution and culling them to approximately agree with
observed colors, \emph{all} available constraints on the target star
are used to condition a direct fit of either a single--, binary--, or
triple--star model to the Dartmouth grid of stellar models
\citep{Dotter:2008, Feiden:2011}.  This fit is done using multi-modal
nested sampling, implemented with \texttt{MultiNest}
\citep{Feroz:2009, Feroz:2011, Feroz:2013}, via the
\texttt{PyMultiNest} wrapper \citep{Buchner:2014}.  Monte Carlo
samples of stellar properties for the population simulations are then
drawn directly from these posterior samples.

As a result, \vespa\ creates full posterior samplings of the physical
properties of the host star, modeled as a single, binary, and triple
star system, as a by-product of the FPP calculation.  Parameters
directly fitted for in this process are stellar mass, age, [Fe/H],
$A_V$ extinction, and distance.  For binary and triple fits, secondary
and/or tertiary mass parameters are added, with all other parameters
assumed to be the same among all components.  Photometric observations
upon which these fits are conditioned are assumed to be the sum of all
components.  If spectroscopic and/or asteroseismic measurements are
used (e.g., constraints on effective temperature or stellar surface
gravity), they are assumed to relate to only the primary star.  Priors
used in these fits are listed in \tab{priors}---notably, we
use a prior on [Fe/H] based on a double-Gaussian fit to the local
metallicity distribution \citep{Hayden:2015, Casagrande:2011}.  Posterior
chains of all other stellar parameters of interest (e.g., temperature,
surface gravity, radius, etc.) are derived from the chains of fitting
parameters by evaluting the stellar models using \isochrones.

\begin{deluxetable}{cc}
\tablewidth{0pt}
\tabletypesize{\scriptsize}
\tablecaption{Priors used in stellar property fits
\tablabel{priors}}
\tablehead{
\colhead{Parameter} &
\colhead{Prior}}
\startdata
Primary mass $M_A$ & $\propto M_A^{-2.35},~M_A > 0.1$ \\
Secondary mass $M_B$ & $\propto (M_B/M_A)^{0.3},~0.1 <= M_B < M_A$ \\
Tertiary mass $M_C$ & $\propto (M_C/M_A)^{0.3},~0.1 <= M_C < M_B$ \\
Age {[}Gyr{]} & $\mathcal U(1,15)$ \tablenotemark{a}\\
{[}Fe/H{]} & $\frac{0.8}{0.15} \mathcal N(0.016, 0.15) + \frac{0.2}{0.22} \mathcal N(-0.15, 0.22)$  \tablenotemark{b} \\
$A_V$ {[}mag{]} & $\mathcal U(0, A_{V, \mathrm{max}})$ \tablenotemark{c} \\
Distance $d$ & $\propto d^2$ 
\enddata
\tablenotetext{a}{The age range for the Dartmouth stellar model grids used.}
\tablenotetext{b}{Double-Gaussian fit to measured local stellar
  metallicity distribution \citep{Hayden:2015, Casagrande:2011}.}
\tablenotetext{c}{Maximum allowed value is the Galactic extinction at
  infinity calculated along the star's line of sight, according to \citet{Schlegel:1998}.}
\end{deluxetable}


\section{Data and Constraints}
\sectlabel{data}

The goal of this work is to calculate the FPP for every KOI,
regardless of classification as CONFIRMED, CANDIDATE, or FALSE
POSITIVE.  As such, we begin with a list of \ntotal\ KOIs from the
Q1-Q17 DR24 table at the NASA Exoplanet Science Institute (NExScI)
Exoplanet Archive (the most recent available uniform catalog).  We
then gather ancillary data and constraints from various sources in
order to enable the \vespa\ calculation:

\begin{enumerate}
\item The RA/Dec coordinates of each star from the Kepler Input
  Catalog (KIC).
\item $grizJHK$ photometry from the KIC, with $griz$ bands corrected
  to the Sloan Digital Sky Survey (SDSS) photometric scale according
  to \citet{Pinsonneault:2012}.
\item Stellar $T_\mathrm{eff}$, [Fe/H], and $\log g$ values and
  uncertainties from the \citet{Huber:2014} stellar properties
  catalog, if the provenance of these values is from spectroscopy or
  asteroseismology.
\item Detrended \kepler\ photometry used for the MCMC modeling of
  \citet{Rowe:2015}, along with information about individually fitted
  transit times, where available.  
\item Best-fit $R_p/R_\star$ from the \citet{Rowe:2015} MCMC analysis.
\item Centroid uncertainty information from the NExScI Exoplanet
  Archive: we assume that the allowed ``exclusion'' radius for a blend
  scenario is 3$\times$ the uncertainty in the fitted centroid
  position (the \verb|koi_dicco_msky_err| column in the Archive
  table).  We floor this value at 0\farcs5, to prevent unrealistically
  small exclusion radii.  If this quantity is not available from the
  Archive we set a default exclusion radius of 4\arcsec.
\item The maximum secondary eclipse depth allowed by the
  \kepler\ photometry.  This quantity is derived by searching the
  phased-folded KOI light curve for the deepest signal at any other
  phase other than that of the primary transit.  This ``model-shift
  uniqueness test'' is described in both Section 3.2.2 of
  \citet{Rowe:2015} and \citet{Coughlin:KSCI}, and the values of these
  metrics for the Q1-Q17 DR24 release will soon be
  published (Coughlin et al.~2015, submitted).  The maximum secondary
  depth we use is
\begin{equation}
\eqlabel{secthresh}
\delta_{\rm max} = \delta_{\rm sec}  + 3 \sigma_{\rm sec},
\end{equation}
where $\delta_{\rm sec}$ is the fitted depth and $\sigma_{\rm sec}$ is
the uncertainty on that depth (including red noise).  As the DR24
pipeline uses two different detrending methods, we calculate
$\delta_{\rm max}$ and $\sigma_{\rm max}$ using both methods and take
the maximum between the two.  When these metrics are not available for
a particular KOI, we default to 10$\times$ the uncertainty in the
\kepler\ pipeline measured transit depth (\verb|koi_depth_err1|).
  
\end{enumerate}
As explained in the \vespa\ documentation, we first specify this
ancillary data in a \verb|star.ini| and \verb|fpp.ini| file for each
KOI, and then for each we run the command-line script \verb|calcfpp|.
This end-to-end calculation (which includes the \isochrones\ fits for
single-, double-, and triple-star models) takes approximately 30
minutes per KOI on a single core, allowing the entire set to be
calculated in approximately one day on the Princeton Univeristy
``Tigress'' computing cluster, using 200 cores.



\section{Results}
\sectlabel{results}

The results of the \vespa\ calculations are presented in
\tab{stars} and \tab{fpp}, and are discussed in
the following subsections.


\subsection{Stellar Properties}
\sectlabel{results:stars}

As discussed in \sect{methods:stellar}, \vespa\ fits for stellar
properties as part of its FPP--calculating procedure, using the
\isochrones\ package.  Thus, we obtain posterior samplings of the
physical properties of each KOI as a side effect of this batch
calculation, a result of general interest independent of FPP.
\tab{stars} presents summarized results of these single--star
fits.  While \vespa\ also fits double-- and triple--star models for
each KOI, these are of less general interest and so we do not
present them separately.

\fig{starsteff} and \fig{starsfehradius} compare the estimated
effective temperatures, metallicities, and radii derived in this work
to those independently determined for (or compiled by) the official
\kepler\ stellar properties catalog \citep[][hereafter
  \citetalias{Huber:2014}]{Huber:2014}.  While there is largely
general agreement, there are also some discrepancies, highlighting some
difficulties of estimating physical stellar properties.

In particular, we note that for stars which \citetalias{Huber:2014}
list as $T_{\rm eff} < 4000$\,K, \isochrones\ predicts systematically
hotter temperatures.  Many of the \citetalias{Huber:2014} properties
for these stars are taken from
\citet[][\citetalias{Dressing:2013}]{Dressing:2013}.  Those properties
were determined by trying to match the $grizJHK$ photometry
of a grid of model stars from the Dartmouth models, supplemented by
some interpolation.  \citetalias{Dressing:2013} also imposed priors
on [Fe/H] and the height of stars above the plane of the Galaxy.  To
validate their methodology, \citetalias{Dressing:2013} compare their
results for 26 nearby stars to the masses predicted for those stars by
combining parallax measurements with the \citet{Delfosse:2000}
relation between mass and absolute $K$-band magnitude.  While they
find general good agreement, \citetalias{Dressing:2013} does note that
their masses are on average about 5\% lower than the Delfosse-predicted
masses.  Our estimated masses for these stars are typically
$\sim$10-15\% higher than those estimated by \citetalias{Dressing:2013}.

The same data ($grizJHK$ photometry) and the same stellar models
(Dartmouth) were used for both this work and
\citetalias{Dressing:2013}, raising the question of the origin of the
systematic differences between these methods.  The primary origin of
this discrepancy appears to be the fact that \isochrones\ performs a
full multi-modal posterior exploration of the stellar parameter space,
marginalizing over the unknown $A_V$ extinction in the process, while
\citetalias{Dressing:2013} uses rather a fixed 1\,mag of $V$-band
extinction per 1000\,pc and selects the maximum-likelihood
match to the grid of models.  As we allow for a maximum extinction up
to the measured $A_V$ extinction at infinity, not explicitly tied to
distance, this typically allows for slightly hotter stars with
slightly more extinction than was permitted by \citet{Dressing:2013}.

The other significant discrepancy between the \isochrones\ results and
\citetalias{Huber:2014} is among evolved stars, as seen in the lower
panel of \fig{starsfehradius}.  Many of these stars have densities
measured via asteroseismology, and \isochrones\ does not unambiguously
identify all of them as evolved.  However, it should be noted that we
do in fact identify over half of them as probably significantly
evolved---this is made possible by the multi-modal posterior sampling
of \texttt{MultiNest} used by \isochrones.  In addition, as the middle
panel of \fig{starsfehradius} shows, even for stars not positively
identified as evolved by \citetalias{Huber:2014}, isochrones often
allows for a significant range of stellar radius---also desirable
behavior, as \citetalias{Huber:2014} estimates the properties for many
of these using only broad-band photometry as well, which means their
true nature is not securely known.  The need for caution when
estimating the radii of KOI host stars has also been emphasized by
\citet{Bastien:2014}, who find from photometric ``flicker''
measurements that a significant number of FGK KOI hosts may be
slightly evolved.

We emphasize that stellar parameter estimation is not the central goal
of this work, nor are the FPP results very sensitive to the exact
estimated stellar properties.  The exception to this would be if the
stellar density estimate is significantly mis-estimated, which would
be the case if a star is not properly identified as evolved.  However,
we note that of the 730 KOIs with host stars $>$2\,$R_{\odot}$ at the
NExScI Archive, 502 of them already have FALSE POSITIVE designations;
additionally, \citet{Sliski:2014} find a large false positive rate for
KOIs with evolved host stars.  Given all of these considerations, we
believe that potential systematic issues with stellar property
determinations in various corners of parameter space do not strongly
affect the main results, which are the astrophysical false positive
probabilities of thousands of KOIs.


\subsection{False Positive Probabilities}
\sectlabel{results:fpp}

Of the \ntotal\ KOIs in the Q1-Q17 DR24 table at the NExScI Exoplanet
Archive, \vespa\ successfully calculates the FPP for \ncalc.  
\Sect{failures} contains detailed explanations of the failure reasons for 
the \nfail\ KOIs for which we do not present \vespa\ results.  
FPPs and their uncertainties are determined as the mean and standard 
deviations of 10 bootstrap 
recalculations of the initially simulated populations for each KOI (as described in 
\sect{methods:fpp}).  These results are listed in \tab{fpp}.  
The median fractional FPP uncertainty 
for KOIs with 0.001 $<$ FPP $<$ 0.1 (within an order of magnitude of the 
validation threshold) is about 12\%, and this distribution 
is shown in \fig{sigmafpp}.

In order to properly interpret these results, it is necessary to
understand the range of applicability of the \vespa\ calculation.
First of all, this method selects between different specific explanations 
for the transit-like signal, and cannot
comment on whether the signal might be caused by stellar variability
or an instrumental false alarm.  Thus, \vespa\ results on low
signal-to-noise ratio (SNR) candidates that are not clearly
transit-like must be viewed with caution.  This being said, the reason 
for including the artificial 
``boxy'' and ``long'' models in the model selection calculation
 (\sect{methods:fpp}) is to flag signals that do not 
fit well with any of the astrophysical eclipse models---in fact, the ``long''
model is preferred ($>50\%$) by \nlongfp\ KOIs that are already dispositioned
as FALSE POSITIVE.

Additionally, an important constraint used in the FPP calculation is the
allowed sky area inside which a blended false positive may live.  As
described in \sect{data}, this value is taken to be three times the
uncertainty on the fitted centroid position from the pixel-level data.
However, many KOIs have already been identified to be blended binary
false positives displaced from the target star.  Some of these are
found by detecting significant centroid offsets in the pixel-level
data---in these cases, \vespa\ treating the confusion radius simply as
the uncertainty in the centroid position will clearly give a
misleading result.  Other displaced false positives have been
identified as originating from displaced stars by finding KOIs with
matching periods and epochs \citep{Coughlin:2014}, and often the
``parents'' of these signals are outside the pixel masks, and so
unable to be detected via centroid-measuring methods.  In these cases
as well, the \vespa\ assumptions break down, and the FPP calculations
will not be valid.

\begin{deluxetable}{cccc}
\tabletypesize{\scriptsize}
\tablecaption{Mean FPPs of candidate KOIs with reliable \vespa\ calculations
\tablabel{meanFPP} }
\tablehead{\colhead{Selection} & \colhead{Number} & \colhead{Mean FPP} } 
\startdata
all & 2857 & 0.155 \\
singles & 1688 & 0.206 \\
multis & 1169 & 0.082 \\
$R_p > 15\,R_\oplus$ & 256 & 0.837 \\
$10\,R_\oplus < R_p < 15\,R_\oplus$ & 91 & 0.220 \\
$4\,R_\oplus < R_p < 10\,R_\oplus$ & 252 & 0.218 \\
$2\,R_\oplus < R_p < 4\,R_\oplus$ & 1160 & 0.066 \\
$R_p < 2\,R_\oplus$ & 1098 & 0.071 
\enddata
\end{deluxetable}

To summarize, the results presented in \tab{fpp} are strictly reliable
only for KOIs \emph{that have already strongly passed the other Kepler
vetting tests, and that are not indicated to be clearly poor fits to
all the proposed hypotheses}.  The first cut for this is the KOI
disposition: FALSE POSITIVE indicates failure of one or more of these
tests (and thus probable invalidity of the \vespa\ calculation).
However, because of the generally permissive philosophy of the
\kepler\ dispositioning, not all CANDIDATE KOIs have the same level of 
reliability in their disposition, due an ``innocent 
until proven guilty'' philosophy.
That is, something is not identified as a FALSE POSITIVE
\emph{unless there is positive confirmation of false positive status}.
In particular, when pixel-level analysis fails to determine whether
the signal is indeed coming from the target star because of low SNR,
such a KOI will still receive CANDIDATE disposition.  Therefore, the
\vespa-calculated FPP may be considered most reliable only when a KOI
is designated a CANDIDATE (or CONFIRMED) and has large enough SNR to
enable secure positional determination.  In addition, for greatest
reliability we also require a signal's multiple-event statistic (MES;
equivalend to SNR) to be greater than 10, in order to avoid low-SNR
signals that might be caused by light-curve systematics.

To enable interpretation, \tab{fpp} thus contains the current KOI
disposition, MES, and the results of the positional probability calculations
of \citet{Bryson:KSCI}.  \Fig{fppall} shows FPPs for all KOIs passing
the following criteria:
\begin{itemize}
\item Dispositioned CANDIDATE or CONFIRMED at the NExScI archive,
\item MES $>$ 10, indicating the signal is unlikely to be caused by 
      systematic noise in the light curve, and
\item Probability $> 0.99$ to be on the target star, according
  to \citet{Bryson:KSCI}, along with a positional probability
  ``score'' $> \posprobthresh$ (indicating a reliable result).  
\end{itemize}

These selections leave \nreliable\ KOIs for which the \vespa\ results
can be considered most reliable;  \Tab{meanFPP} presents the mean FPPs
for different subsets, showing several notable features.  Single KOIs
are about 2.5$\times$ more likely to be false positives than KOIs in
multiple-KOI systems, in qualitative agreement with
\citet{Lissauer:2012}---and this is true even without giving any
``multiplicity boost'' to multi-KOI systems for the increased transit
probability of subsequent planets once one planet transits in a
coplanar system.  Also, large candidates typically have high FPPs, in
agreement with \citet{Santerne:2012} and \citet{Santerne:2015}.  And
finally, the mean FPPs are very consistent with the \textit{a priori}
predictions of \citet{Morton:2011b} and with the analysis of
\citet{Fressin:2013}.

\subsection{Unidentified Ephemeris Matches?}
\sectlabel{PEMs}

One potential concern worth addressing in some detail, before deciding
which planets to validate based on the \vespa\ calculations, is the 
possibility of false positives caused by distant contamination
but not identified through the ``period-epoch match'' (PEM) technique
used by \citet{Coughlin:2014}, due to the fact that not all stars in the 
\kepler\ field were monitored by the mission.  We thus estimate here
the probability that a CANDIDATE KOI to which \vespa\ assigns a low FPP
might still be caused by such a scenario.

In the Q1-Q17 DR24 KOI table used as the basis for this work, 980
KOIs were identified as PEMs.  Of these, 187 were not identified 
as false positives by any other method.  Only 15 of these 187 survived 
all the quality cuts described in \sect{results:fpp}. And of these 15, 
only 3 have FPP $< 0.01$.  (See \sect{gaidos} for an example of a KOI caused by a 
``column anomaly'' effect that went unidentified by the \kepler\ pipeline but
to which \vespa\ assigned a high FPP.)  Thus, we expect only about 0.3\% (3/980) of 
as-yet unidentified distant-contamination FPs to end
up with FPP $< 0.01$ according to \vespa. As the fraction of false positives
from pixel contamination but unidentified as PEMs among the entire KOI sample is
estimated to be something around 23\% \citep{Coughlin:2014}, we estimate
that there remains a small ($\sim$0.06\%) residual FPP for all KOIs, even 
when the \vespa-calculated FPP is negligibly tiny.

\subsection{Validation of \nvalnew\ new planets}

While the FPP below which to claim planet validation is clearly
an arbitrary choice, there is precedent to using ${\rm FPP} <0.01$ as 
the threshold---\citet{Rowe:2014} validated over 800 multi-planet KOIs
using this number, and \citet{Montet:2015} used it to validate a sample
of K2-Campaign 1 planets.  Adopting this same threshold and adding
the 0.06\% residual FPP estimated in \sect{PEMs} to the \vespa-calculated
value, we find that 
of the KOIs with reliable \vespa\ FPPs, \nval\ have ${\rm FPP}
<0.01$, and are thus validated at the 99\% level.  
These KOIs are labeled as such in \tab{fpp}.  These are not all new validations,
however, since a number of them are already CONFIRMED. \Fig{fpppie}
shows a different kind of summary, grouping KOIs by disposition and
splitting up the CANDIDATES according to the \vespa\ results, showing
that \nvalnew\ KOIs are newly validated at the 99\% level.
\Fig{rpcand} shows the radii and periods of the CONFIRMED and 
CANDIDATE KOIs, with the transparency of the points representing
the \vespa-calculated FPP.  \Fig{validatedcompare} compares the temperature
and radii of the stars hosting these planets between the \citetalias{Huber:2014}
and \isochrones\ analysis.

\begin{deluxetable*}{llccccccc}
\tabletypesize{\scriptsize}
\tablewidth{0pc}
\tablecaption{Newly Validated Planets in the Optimistic Habitable Zone.\tablabel{hz}}
\tablehead{
\colhead{KOI} &
\colhead{Kepler name} &
\colhead{Period} &
\colhead{\rpl} &
\colhead{$F_{\rm p}$} &
\colhead{\teff} &
\colhead{\logg} &
\colhead{\rstar} &
\colhead{\mstar} \\
\colhead{} &
\colhead{} &
\colhead{[days]} &
\colhead{\rearth} &
\colhead{$F_{\rm e}$} &
\colhead{[K]} &
\colhead{} &
\colhead{\rsun} &
\colhead{\msun}
}
\startdata
463.01 &   Kepler-560 b   & 18.478 & 1.57$^{+0.26}_{-0.26}$ & 1.26$^{+0.58}_{-0.43}$ & 3387$^{+59}_{-50}$ & 4.96$^{+0.10}_{-0.10}$ & 0.30$^{+0.05}_{-0.05}$ & 0.30$^{+0.05}_{-0.05}$ \\
854.01 &   Kepler-705 b   & 56.056 & 1.96$^{+0.25}_{-0.25}$ & 0.64$^{+0.23}_{-0.18}$ & 3593$^{+58}_{-66}$ & 4.78$^{+0.08}_{-0.08}$ & 0.47$^{+0.06}_{-0.06}$ & 0.49$^{+0.06}_{-0.06}$ \\
2418.01 &   Kepler-1229 b   & 86.829 & 1.12$^{+0.13}_{-0.22}$ & 0.35$^{+0.12}_{-0.14}$ & 3724$^{+60}_{-74}$ & 4.84$^{+0.09}_{-0.08}$ & 0.41$^{+0.05}_{-0.08}$ & 0.43$^{+0.05}_{-0.07}$ \\
3010.01 &   Kepler-1410 b   & 60.866 & 1.56$^{+0.15}_{-0.15}$ & 0.93$^{+0.25}_{-0.21}$ & 3903$^{+50}_{-60}$ & 4.74$^{+0.06}_{-0.06}$ & 0.52$^{+0.05}_{-0.05}$ & 0.54$^{+0.05}_{-0.05}$ \\
3282.01 &   Kepler-1455 b   & 49.277 & 1.97$^{+0.25}_{-0.19}$ & 1.30$^{+0.50}_{-0.34}$ & 3894$^{+83}_{-101}$ & 4.71$^{+0.06}_{-0.07}$ & 0.54$^{+0.07}_{-0.05}$ & 0.55$^{+0.06}_{-0.06}$ \\
4036.01 &   Kepler-1544 b   & 168.811 & 1.83$^{+4.73}_{-0.17}$ & 1.02$^{+13.70}_{-0.25}$ & 4893$^{+141}_{-110}$ & 4.54$^{+0.06}_{-0.98}$ & 0.76$^{+1.97}_{-0.07}$ & 0.73$^{+0.28}_{-0.06}$ \\
4356.01 &   Kepler-1593 b   & 174.510 & 1.91$^{+0.16}_{-0.21}$ & 0.29$^{+0.09}_{-0.09}$ & 4366$^{+131}_{-166}$ & 4.82$^{+0.06}_{-0.05}$ & 0.46$^{+0.04}_{-0.05}$ & 0.49$^{+0.03}_{-0.05}$ \\
4450.01 &   Kepler-1606 b   & 196.435 & 1.98$^{+0.72}_{-0.15}$ & 1.38$^{+1.48}_{-0.32}$ & 5536$^{+161}_{-140}$ & 4.57$^{+0.03}_{-0.24}$ & 0.82$^{+0.29}_{-0.06}$ & 0.90$^{+0.09}_{-0.09}$ \\
5856.01 &   Kepler-1638 b   & 259.337 & 1.70$^{+0.76}_{-0.21}$ & 1.47$^{+2.03}_{-0.44}$ & 5906$^{+183}_{-147}$ & 4.47$^{+0.10}_{-0.29}$ & 0.85$^{+0.38}_{-0.10}$ & 0.77$^{+0.10}_{-0.04}$ \\
\enddata
\tablecomments{This table lists CANDIDATE KOIs validated in this work that may lie within the optimistic habitable zones of their host stars.  The stellar and planetary properties for this table are taken from the DR24 table at the NExScI Exoplanet Archive.  Further individual
study of each of these systems using detailed follow-up observations will either solidify or amend their potentially habitable nature.  In particular, we note that high-resolution imaging observations on the CFOP archive\footnote{\url{http://cfop.ipac.caltech.edu}} reveal both KOI-2418 and KOI-3010 to have close companions which may or may not affect their habitable nature.}
\end{deluxetable*}

As there is significant interest in identifying potentially habitable
planets, \tab{hz} lists the properties of \nhz\ CANDIDATE KOIs newly
validated by this work that \textit{may} fall within the optimistic
habitable zones of their host stars \citep{Kopparapu:2013}, 
according to the stellar
properties reported by the DR24 table at the NExScI Archive.  We note
that while more detailed follow-up observations (imaging and
spectroscopy) have been taken for each of these targets, we make no
attempt here to characterize these systems in detail.  However, 
we do note that high-resolution imaging of KOI-2418 and KOI-3010 reveal
that these two host stars have close companions that may or may not
change the habitable nature of the planets (e.g., if the planets happen
to transit the secondary star instead of the primary).  Additionally, analysis of
high-resolution spectroscopy will solidify the
properties of all these host stars, affecting the habitable zone
boundaries.  We thus emphasize that this list is neither complete nor
final, serving only to draw attention to new validations of 
interest rather than to be a definitive statement on potential 
habitability.  We also note that based solely on the properties in the 
DR24 table, KOI-5475.01 (listed as having a 448-day orbital period) 
should also be included in this list of potentially habitable-zone
planets.  However, as explained in \citet[][Section 5.5.4]{Coughlin:2015},
this particular KOI actually has a 224-day period, making its insolation
too high to be within the habitable zone; we have thus excluded it 
from \tab{hz}.

\subsection{Likely False Positives}

In addition to the confident validations, we identify \nfpnew\ KOIs
that currently have CANDIDATE disposition but for which \vespa\
calculates FPP $>$ 0.9; these KOIs are likely to be false positives.
As \Fig{rpcand} shows, many of these newly identified false positives
have large radii---this is again because of the dispositioning
philosophy adopted by the \kepler\ team, which does not use any cut in
transit depth or inferred ``planet'' size to identify FALSE POSITIVES.
To identify these likely false positives from the \vespa\
calculations, we do not require CANDIDATES to obey the same selections
we used to ensure a clean sample for validation.  This is because even
if a signal has characteristics such that a low \vespa\ FPP would not
be sufficient to validate it, a high \vespa\ FPP is still sufficient
to cast doubt on its planetary nature. As a demonstration of the
ability of \vespa\ to identify false positives, \Fig{fppfps} shows the
FPPs for KOIs that are dispositioned FALSE POSITIVE and have MES $>$
10. The vast majority of these have large FPPs.

One final note about these calculations is that unidentified transit
timing variations (TTVs) will increase the FPP of a transit signal, as
the shape of the folded light curve will be distorted, typically
resulting in a longer signal which a trapezoid model will also
identify as more V-shaped.  While we have analyzed light curves
correcting for known TTVs when available, we have also undoubtedly
missed many systems with as-yet-unspecified TTV signals.  This means
that a large FPP may be simply an indication of unidentified TTVs
rather than an astrophysical false positive, especially in  
multi-planet systems, which should overall have a very low false positive
rate \citep{Lissauer:2014,Rowe:2014}.  It is thus probable that
despite multi-KOIs having lower FPPs than singles (\tab{meanFPP}),
even these relatively low FPPs are inflated by the effect of TTVs.
\Tab{fpp} lists whether known TTVs were accounted for when constructing
the folded transit light curve for each KOI.


\subsection{Blended Transiting Planets?}
\sectlabel{btp}

In this work, the only astrophysical false positive scenarios we consider are 
eclisping binary stars (EB, HEB, BEB, and the double-period versions
thereof).  Previous work studying \kepler\ false positive rates
\citep[e.g.][]{Fressin:2013} has also considered the  ``blended transiting
planet'' to be a false positive---i.e.,  a fainter companion
star hosting a transiting planet larger than what would be inferred if
it were transiting the target star.  Because we do not consider such
a scenario to be a false positive, the \vespa\ analysis presented here 
does not quantify its probability.  As a result, we are \textit{not}
able to unambiguously determine the radii of the planets we validate---
all the planet radii listed in \tab{fpp} are based on the assumption
that the planet transits the target star and that the target star is 
unblended.  If the target star is actually a member of an unresolved
binary system (as a significant fraction of KOIs undoubtedly are),
then the true planet radius will be larger (significantly larger 
if transiting a fainter companion).  This was indeed the motivation
for \citet{Fressin:2013} to consider the ``blended transiting planet''
as a false positive; part of the goal of that work was to 
compute the planet occurrence rates in different radius bins.  
However, trying to distinguish between single and binary target
star configurations is beyond the scope of this work; we thus follow
the precedent of \citet[][especially Section 5]{Lissauer:2014} by 
acknowledging the potential for substantial radius
uncertainties among the validated planet sample while nevertheless 
defending the validations themselves as robust.

\subsection{Failure Modes}
\sectlabel{failures}

We were not able to successfully run \vespa\ on all of the KOIs. The 
various reasons for these occasional failures are detailed below.  
\begin{enumerate}
\item \nbadphot\ KOIs did not receive MCMC modeling.  Most of these
  have been are already designated FALSE POSITIVE at the archive.
\item \nbadrowemcmc\ KOIs did receive MCMC modeling but had unphysical fit
  results; e.g., negative $R_p/R_\star$ or best-fit impact parameter
  greater than $(1 + R_p/R_\star)$.  These KOIs were left out of the 
  \vespa\ calculations.
\item The host stars of \nbadstellar\ KOIs are not included in the
  \citetalias{Huber:2014} stellar property catalog, and thus were not
  included in this analysis.
\item \nbadsec\ KOIs have no \verb|koi_depth_err1| value on the archive,
  and thus had no weak secondary constraint and were left out of the
  calculations.
\item For \nbadtrapfit\ KOIs, the trapezoid MCMC fit did not converge.  The
  convergence criterion was for the autocorrelation time of the chain
  for each parameter to be shorter than 10\% of the total chain
  length.
\item For \nbadroche\ KOIs, the orbital period and stellar properties of the 
  candidate imply the orbit to be within its host star's Roche limit.  This
  usually happens when the host star is estimated to be a giant, and
  these situations are nearly always false positives. 
\end{enumerate}
The numbers in this list correspond to the numbers in
the ``failure'' column in \tab{fpp}.  

\section{Comparison with follow-up observational studies}
\sectlabel{comparisons} 

One of the difficulties with probabilistic
validation is that, by necessity, it is typically invoked when no
other method of follow-up confirmation is possible.  It can therefore
be difficult to find ways to compare the results of a calculation such
as \vespa\ to any known observational ground-truth, or to ``validate
the validations.''  However, because so much follow-up observational
effort has gone into \kepler\ candidates over the last few years,
there actually are two different datasets that do provide such
information for relatively small subsets of candidates.  Here, we
discuss the results of \vespa\ calculations for these candidates and
the impliciations for the reliability of the \vespa\ framework.

\subsection{\textit{Spitzer} photometry}
 
The first of these datasets is from \citet{Desert:2015}, who observe
the transits of 50 \kepler\ candidates with the \textit{Spitzer} Space
Telescope in order to observationally constrain the FP rate.  The idea
behind these observations is that a blended EB false positive will
often show a color-dependent transit depth, and so comparing the
candidate depths measured by \kepler\ to those measured in the
infrared by Spitzer would give an idea of how likely a signal is to be
a false positive. The results of their analysis suggest that fewer
than 8\% of their observed candidates are likely to be false
positives.  Of these 50 candidates, \vespa\ on its own calculates FPP
$>$ 0.1 for four.  Two of these (KOI-103.01 and KOI-248.02) are
systems with known significant TTVs; the other two (KOI-247.01 and
KOI-555.02) are likely false positives, with FPPs of 0.90 and 0.81,
respectively.  For all but two of the remaining candidates, \vespa\
gives FPP $< 0.01$.  The false positive rates calculated in
\citet{Desert:2015} are also based on probabilistic arguments very
similar to \citet{Morton:2011b} and thus are not quite candidates for
``ground-truth'' comparison, but the lack of transit chromaticity and
the agreements between these two independent studies are certainly
supportive of the \vespa\ results.

\subsection{Radial-velocity monitoring of large candidates}

A much more independent and powerful test data set has recently become
available in the work of \citet{Santerne:2015}, which presented the
results of a long-term RV-monitoring campaign targeting 129 \kepler\
giant-planet candidates.  Of these, they confirm 45 to be planets and
identify 48 as eclipsing binaries, 15 as contaminating EBs (CEBs), and
3 as brown dwarfs.  They are unable to determine the nature of the
remaining 18.  These results imply a relatively high FP rate among
giant-planet candidates, possibly near 50\%, which sounds potentially
concerning, although we reiterate that KOIs are not ruled FALSE 
POSITIVES based on their transit depth or inferred size alone.
Additionally, when we look at the \vespa\ results on this sample, we
see very good agreement between our results and the RV-
detected ``ground truth'' (\Tab{santerne}).  Notably, confirmed
planets show a mean FPP of about 10\% and a median of much less than 1\%,
while confirmed EBs (CEBs) show a mean FPP of 75\% (78\%) and a median
of 97\% (99\%).  The three brown dwarfs also show low FPPs, which is
understandable because \vespa\ doesn't pretend to predict anything
about the mass of the companion, and BDs are essentially the same size
as giant planets.  It is also instructive to investigate the four
cases where \vespa\ computes high FPPs ($>$0.5) for confirmed planets.
Three of these (KOI-377.01, KOI-1426.02 and KOI-1474.01) have
significant TTVs, and one is a grazing eclipse (KOI-614.01).
KOI-1474.01 is also on a highly eccentric orbit \citep{Dawson:2012},
in addition to its TTVs, which also contributes to its high FPP.
Overall, this comparison powerfully demonstrates the reliability of
the \vespa\ calculation, showing that even in a population of
candidates that include many false positives, it is able to
effectively identify which are true planets and which are not.

\subsection{The enigmatic case of KOI-6705.01}
\sectlabel{gaidos}

We also briefly discuss another case of individual interest.
KOI-6705.01, a 0.99\,d signal around a mid M-dwarf star, was
identified  as a KOI of possible interest by \citet{Gaidos:2015}.
After significant follow-up observations and detailed full-frame image
analysis, they concluded that the signal was most likely due to a
charge-transfer  effect from a 1.99\,d EB located on the same CCD
column.  For this KOI, \vespa\ calculates an FPP of 1, with the 
``long'' model preferred by far (of the astrophysical models, the 
double-period BEB---the true scenario---is preferred). This is an excellent
example of what was discussed in \sect{PEMs}: that even effects like
column anomalies that were not identified by the \kepler\ team as
ephemeris matches will typically be identified as false positives by 
\vespa.

\begin{deluxetable}{cccc}
\tabletypesize{\scriptsize}
\tablecaption{\vespa--calculated FPPs of the \\Santerne (2015) RV sample
  \tablabel{santerne}}
\tablehead{\colhead{RV-based nature} & \colhead{Number} & \colhead{mean FPP} & \colhead{median FPP}}
\startdata
Planet & 43 & 0.1 & 3.6e-05 \\ 
Brown dwarf & 3 & 0.012 & 0.0026 \\ 
Eclipsing binary (EB) & 43 & 0.75 & 0.97 \\ 
Contaminating EB & 13 & 0.78 & 0.99 \\ 
Undetermined & 18 & 0.31 & 0.01 
\enddata
\end{deluxetable}


\section{Conclusions}
\sectlabel{conclusions}

In this work, we have calculated the astrophysical false positive
probability (FPP) for every \kepler\ object of interest (KOI) in the
Q1-Q17 DR24 table, using the publicly available Python module
\vespa, which implements the procedure introduced in
\citet{Morton:2012}, with improvements in stellar parameter modeling
and the inclusion of new ``double-period'' false positive scenarios 
and artificial models to identify KOIs that are not well explained
by any of the astrophysical models.  We have also for the first time
estimated uncertainties in the \vespa\ calculation, through a bootstrap
resampling procedure.

While the assumptions behind this calculation are not necessarily
valid for every KOI (see \sect{results:fpp}), we have identified
\nvalnew\ KOIs that have reliable FPPs of $< 0.01$, resulting in
validation of their planetary nature at the 99\% confidence level,
more than doubling the number of confirmed \kepler\ exoplanets.  Among
this set of newly validated planets are \nhz\ that are consistent with being
in the habitable zones of their host stars.  We also identify \nfpnew\ new likely
false positive KOIs, although we note that some of these may be due to
unidentified or miscorrected transit timing variations.  

The reliability of these calculations depends significantly on the
results of \citet{Bryson:KSCI}, which quantify the probability that
the eclipse/transit signal is spatially coincident on the sky with the
presumed target star.  Without confirmation that the transit signal is
not coming from a significantly displaced source, the sky area used as
part of the prior for the false positive scenarios would need to be
significantly larger than the positional uncertainty value assumed in
this work (described in \sect{methods:fpp}).  Additionally, the
blended false positive scenarios \vespa\ considers are assumed to be
fully contained within the photometric aperture; this assumption would
also be broken if the source of the eclipse were significantly
displaced.  We estimate that perhaps 0.06\% of the planets we validate
could be signals coming from significantly displaced sources, similar 
to those identified as period-epoch matches by \citet{Coughlin:2014}
but unidentified by that analysis.

While previous \emph{a priori} false positive rate estimates
\citep{Morton:2011b,Fressin:2013} have made clear that the \kepler\
planet candidate catalogs are generally low enough to ignore for the
purposes of planet occurrence rate calculations, any more detailed
study of any small subset of individual KOIs should understand in more
detail the FPPs of those specific candidates.  This type of small-sample 
candidate culling using individually calculated FPPs has
already been done in the literature
\citep{MortonSwift:2014,MortonWinn:2014}; the publication of this full
catalog allows the community to do the same.  In particular, several
other studies have shown that specific samples of KOIs tend to have
larger FPPs than the global average, so studies involving giant-planet
KOIs \citep{Santerne:2015} or evolved stars \citep{Sliski:2014} are in
even greater need of the individual FPP analysis here presented.

The \kepler\ mission has demonstrated that space-based transiting
planet surveys identify planet candidates at a rate much faster than
traditional follow-up techniques can confirm them.  As a result, false
positive probability quantification techniques are now an integral
part of the landscape of exoplanet science.  While the present work is
the first large-scale demonstration of a fully automated validation
procedure, there is much progress still to be made.  For example,
there is currently no support within \vespa\ to calculate the FPP for
a candidate which has a \emph{specifically identified} but previously
unknown close companion.  Future development plans for the
\vespa\ package include support for this scenario, as well as other
improvements.  One of the most important of these will be to allow for
a contaminating EB to not be fully contained within the target
photometric aperture; that is, modeling the probability for
further-away stars to be EBs contributing only a small amount of their
flux to the target photometry, such as is the case for many of the
false positives identified via ephemeris matching by
\citet{Coughlin:2014}.  Full inclusion of this effect will allow for
even low-SNR candidates to receive confident \vespa\ analysis, which
is now limited only to KOIs for which confident pixel-level positional
analysis is possible.

Beyond \kepler, future transit missions such as TESS and PLATO will
require automated false positive analysis in order to efficiently sift
through the large numbers of candidates that they will find.  This
work demonstrates that \vespa\ will be a valuable tool towards this
purpose.

\vspace{1.5\baselineskip}
Instructions for reproducing all the calculations presented 
herein can be found at \url{https://github.com/timothydmorton/koi-fpp}.
Summary figures produced by \vespa\ for all KOIs can be accessed
at \url{http://kepler-fpp.space}. 

\acknowledgments TDM acknowledges support from the \kepler\
Participating Scientist Program (NNX14AE11G). We thank the anonymous
referee, Daniel Huber, Jack Lissauer, Juna Kollmeier, and David Hogg 
for providing helpful suggestions that improved both the analysis and 
presentation herein.
This paper includes data collected by the \kepler\ mission. Funding for the \kepler\ mission
is provided by the NASA Science Mission directorate. The authors
acknowledge the efforts of the \kepler\ Mission team for obtaining
the light curve products used in
this publication, which were generated by the \kepler\ Mission
science pipeline through the efforts of the \kepler\ Science
Operations Center and Science Office. The \kepler\ Mission is led by
the project office at NASA Ames Research Center. Ball Aerospace built
the \kepler\ photometer and spacecraft which is operated by the
mission operations center at LASP. These data products are archived at
the NASA Exoplanet Science Institute, which is operated by the
California Institute of Technology, under contract with the National
Aeronautics and Space Administration under the Exoplanet Exploration
Program. This research has made use of NASA's Astrophysics Data
System. 

\clearpage
\bibliography{ms}

\begin{thebibliography}{}
\expandafter\ifx\csname natexlab\endcsname\relax\def\natexlab#1{#1}\fi

\bibitem[{{Bastien} {et~al.}(2014){Bastien}, {Stassun}, \&
  {Pepper}}]{Bastien:2014}
{Bastien}, F.~A., {Stassun}, K.~G., \& {Pepper}, J. 2014, \apjl, 788, L9

\bibitem[{{Becker} {et~al.}(2015){Becker}, {Vanderburg}, {Adams}, {Rappaport},
  \& {Schwengeler}}]{Becker:2015}
{Becker}, J.~C., {Vanderburg}, A., {Adams}, F.~C., {Rappaport}, S.~A., \&
  {Schwengeler}, H.~M. 2015, \apjl, 812, L18

\bibitem[{{Borucki} {et~al.}(2012){Borucki}, {Koch}, {Batalha}, {Bryson},
  {Rowe}, {Fressin}, {Torres}, {Caldwell}, {Christensen-Dalsgaard}, {Cochran},
  {DeVore}, {Gautier}, {Geary}, {Gilliland}, {Gould}, {Howell}, {Jenkins},
  {Latham}, {Lissauer}, {Marcy}, {Sasselov}, {Boss}, {Charbonneau}, {Ciardi},
  {Kaltenegger}, {Doyle}, {Dupree}, {Ford}, {Fortney}, {Holman}, {Steffen},
  {Mullally}, {Still}, {Tarter}, {Ballard}, {Buchhave}, {Carter},
  {Christiansen}, {Demory}, {D{\'e}sert}, {Dressing}, {Endl}, {Fabrycky},
  {Fischer}, {Haas}, {Henze}, {Horch}, {Howard}, {Isaacson}, {Kjeldsen},
  {Johnson}, {Klaus}, {Kolodziejczak}, {Barclay}, {Li}, {Meibom}, {Prsa},
  {Quinn}, {Quintana}, {Robertson}, {Sherry}, {Shporer}, {Tenenbaum},
  {Thompson}, {Twicken}, {Van Cleve}, {Welsh}, {Basu}, {Chaplin}, {Miglio},
  {Kawaler}, {Arentoft}, {Stello}, {Metcalfe}, {Verner}, {Karoff}, {Lundkvist},
  {Lund}, {Handberg}, {Elsworth}, {Hekker}, {Huber}, {Bedding}, \&
  {Rapin}}]{Borucki:2012}
{Borucki}, W.~J., {Koch}, D.~G., {Batalha}, N., {et~al.} 2012, \apj, 745, 120

\bibitem[{{Bryson} \& {Morton}(2015)}]{Bryson:KSCI}
{Bryson}, S.~T., \& {Morton}, T.~D. 2015, KSCI-19092-001

\bibitem[{{Bryson} {et~al.}(2013){Bryson}, {Jenkins}, {Gilliland}, {Twicken},
  {Clarke}, {Rowe}, {Caldwell}, {Batalha}, {Mullally}, {Haas}, \&
  {Tenenbaum}}]{Bryson:2013}
{Bryson}, S.~T., {Jenkins}, J.~M., {Gilliland}, R.~L., {et~al.} 2013, \pasp,
  125, 889

\bibitem[{{Buchner} {et~al.}(2014){Buchner}, {Georgakakis}, {Nandra}, {Hsu},
  {Rangel}, {Brightman}, {Merloni}, {Salvato}, {Donley}, \&
  {Kocevski}}]{Buchner:2014}
{Buchner}, J., {Georgakakis}, A., {Nandra}, K., {et~al.} 2014, \aap, 564, A125

\bibitem[{{Burke} {et~al.}(2015){Burke}, {Christiansen}, {Mullally}, {Seader},
  {Huber}, {Rowe}, {Coughlin}, {Thompson}, {Catanzarite}, {Clarke}, {Morton},
  {Caldwell}, {Bryson}, {Haas}, {Batalha}, {Jenkins}, {Tenenbaum}, {Twicken},
  {Li}, {Quintana}, {Barclay}, {Henze}, {Borucki}, {Howell}, \&
  {Still}}]{Burke:2015}
{Burke}, C.~J., {Christiansen}, J.~L., {Mullally}, F., {et~al.} 2015, \apj,
  809, 8

\bibitem[{{Casagrande} {et~al.}(2011){Casagrande}, {Sch{\"o}nrich}, {Asplund},
  {Cassisi}, {Ram{\'{\i}}rez}, {Mel{\'e}ndez}, {Bensby}, \&
  {Feltzing}}]{Casagrande:2011}
{Casagrande}, L., {Sch{\"o}nrich}, R., {Asplund}, M., {et~al.} 2011, \aap, 530,
  A138

\bibitem[{{Coughlin} {et~al.}(2014){Coughlin}, {Thompson}, {Bryson}, {Burke},
  {Caldwell}, {Christiansen}, {Haas}, {Howell}, {Jenkins}, {Kolodziejczak},
  {Mullally}, \& {Rowe}}]{Coughlin:2014}
{Coughlin}, J.~L., {Thompson}, S.~E., {Bryson}, S.~T., {et~al.} 2014, \aj, 147,
  119

\bibitem[{{Coughlin} {et~al.}(2015{\natexlab{a}}){Coughlin}, {Bryson}, {Burke},
  {Catanzarite}, {Haas}, {Jenkins}, {Li}, {Mullally}, {Rowe}, {Seader},
  {Thompson}, \& {Twicken}}]{Coughlin:KSCI}
{Coughlin}, J.~L., {Bryson}, S.~T., {Burke}, C.~J., {et~al.}
  2015{\natexlab{a}}, KSCI-19104-001

\bibitem[{{Coughlin} {et~al.}(2015{\natexlab{b}}){Coughlin}, {Mullally},
  {Thompson}, {Rowe}, {Burke}, {Latham}, {Batalha}, {Ofir}, {Quarles}, {Henze},
  {Wolfgang}, {Caldwell}, {Bryson}, {Shporer}, {Catanzarite}, {Akeson},
  {Barclay}, {Borucki}, {Boyajian}, {Campbell}, {Christiansen}, {Girouard},
  {Haas}, {Howell}, {Huber}, {Jenkins}, {Li}, {Patil-Sabale}, {Quintana},
  {Ramirez}, {Seader}, {Smith}, {Tenenbaum}, {Twicken}, \&
  {Zamudio}}]{Coughlin:2015}
{Coughlin}, J.~L., {Mullally}, F., {Thompson}, S.~E., {et~al.}
  2015{\natexlab{b}}, ArXiv e-prints, arXiv:1512.06149

\bibitem[{{Dawson} {et~al.}(2012){Dawson}, {Johnson}, {Morton}, {Crepp},
  {Fabrycky}, {Murray-Clay}, \& {Howard}}]{Dawson:2012}
{Dawson}, R.~I., {Johnson}, J.~A., {Morton}, T.~D., {et~al.} 2012, \apj, 761,
  163

\bibitem[{{Delfosse} {et~al.}(2000){Delfosse}, {Forveille}, {S{\'e}gransan},
  {Beuzit}, {Udry}, {Perrier}, \& {Mayor}}]{Delfosse:2000}
{Delfosse}, X., {Forveille}, T., {S{\'e}gransan}, D., {et~al.} 2000, \aap, 364,
  217

\bibitem[{{D{\'e}sert} {et~al.}(2015){D{\'e}sert}, {Charbonneau}, {Torres},
  {Fressin}, {Ballard}, {Bryson}, {Knutson}, {Batalha}, {Borucki}, {Brown},
  {Deming}, {Ford}, {Fortney}, {Gilliland}, {Latham}, \&
  {Seager}}]{Desert:2015}
{D{\'e}sert}, J.-M., {Charbonneau}, D., {Torres}, G., {et~al.} 2015, \apj, 804,
  59

\bibitem[{{D{\'{\i}}az} {et~al.}(2014){D{\'{\i}}az}, {Almenara}, {Santerne},
  {Moutou}, {Lethuillier}, \& {Deleuil}}]{Diaz:2014}
{D{\'{\i}}az}, R.~F., {Almenara}, J.~M., {Santerne}, A., {et~al.} 2014, \mnras,
  441, 983

\bibitem[{{Dotter} {et~al.}(2008){Dotter}, {Chaboyer}, {Jevremovi{\'c}},
  {Kostov}, {Baron}, \& {Ferguson}}]{Dotter:2008}
{Dotter}, A., {Chaboyer}, B., {Jevremovi{\'c}}, D., {et~al.} 2008, \apjs, 178,
  89

\bibitem[{{Dressing} \& {Charbonneau}(2013)}]{Dressing:2013}
{Dressing}, C.~D., \& {Charbonneau}, D. 2013, \apj, 767, 95

\bibitem[{{Fabrycky} {et~al.}(2012){Fabrycky}, {Ford}, {Steffen}, {Rowe},
  {Carter}, {Moorhead}, {Batalha}, {Borucki}, {Bryson}, {Buchhave},
  {Christiansen}, {Ciardi}, {Cochran}, {Endl}, {Fanelli}, {Fischer}, {Fressin},
  {Geary}, {Haas}, {Hall}, {Holman}, {Jenkins}, {Koch}, {Latham}, {Li},
  {Lissauer}, {Lucas}, {Marcy}, {Mazeh}, {McCauliff}, {Quinn}, {Ragozzine},
  {Sasselov}, \& {Shporer}}]{Fabrycky:2012}
{Fabrycky}, D.~C., {Ford}, E.~B., {Steffen}, J.~H., {et~al.} 2012, \apj, 750,
  114

\bibitem[{{Feiden} {et~al.}(2011){Feiden}, {Chaboyer}, \&
  {Dotter}}]{Feiden:2011}
{Feiden}, G.~A., {Chaboyer}, B., \& {Dotter}, A. 2011, \apjl, 740, L25

\bibitem[{{Feroz} {et~al.}(2009){Feroz}, {Hobson}, \& {Bridges}}]{Feroz:2009}
{Feroz}, F., {Hobson}, M.~P., \& {Bridges}, M. 2009, \mnras, 398, 1601

\bibitem[{{Feroz} {et~al.}(2011){Feroz}, {Hobson}, \& {Bridges}}]{Feroz:2011}
---. 2011, {MultiNest: Efficient and Robust Bayesian Inference}, Astrophysics
  Source Code Library, ascl:1109.006

\bibitem[{{Feroz} {et~al.}(2013){Feroz}, {Hobson}, {Cameron}, \&
  {Pettitt}}]{Feroz:2013}
{Feroz}, F., {Hobson}, M.~P., {Cameron}, E., \& {Pettitt}, A.~N. 2013, ArXiv
  e-prints

\bibitem[{{Ford} {et~al.}(2012){Ford}, {Fabrycky}, {Steffen}, {Carter},
  {Fressin}, {Holman}, {Lissauer}, {Moorhead}, {Morehead}, {Ragozzine}, {Rowe},
  {Welsh}, {Allen}, {Batalha}, {Borucki}, {Bryson}, {Buchhave}, {Burke},
  {Caldwell}, {Charbonneau}, {Clarke}, {Cochran}, {D{\'e}sert}, {Endl},
  {Everett}, {Fischer}, {Gautier}, {Gilliland}, {Jenkins}, {Haas}, {Horch},
  {Howell}, {Ibrahim}, {Isaacson}, {Koch}, {Latham}, {Li}, {Lucas}, {MacQueen},
  {Marcy}, {McCauliff}, {Mullally}, {Quinn}, {Quintana}, {Shporer}, {Still},
  {Tenenbaum}, {Thompson}, {Torres}, {Twicken}, {Wohler}, \& {Kepler Science
  Team}}]{Ford:2012}
{Ford}, E.~B., {Fabrycky}, D.~C., {Steffen}, J.~H., {et~al.} 2012, \apj, 750,
  113

\bibitem[{{Foreman-Mackey} {et~al.}(2013){Foreman-Mackey}, {Hogg}, {Lang}, \&
  {Goodman}}]{emcee}
{Foreman-Mackey}, D., {Hogg}, D.~W., {Lang}, D., \& {Goodman}, J. 2013, \pasp,
  125, 306

\bibitem[{{Foreman-Mackey} {et~al.}(2014){Foreman-Mackey}, {Hogg}, \&
  {Morton}}]{DFM:2014}
{Foreman-Mackey}, D., {Hogg}, D.~W., \& {Morton}, T.~D. 2014, \apj, 795, 64

\bibitem[{{Fressin} {et~al.}(2013){Fressin}, {Torres}, {Charbonneau}, {Bryson},
  {Christiansen}, {Dressing}, {Jenkins}, {Walkowicz}, \&
  {Batalha}}]{Fressin:2013}
{Fressin}, F., {Torres}, G., {Charbonneau}, D., {et~al.} 2013, \apj, 766, 81

\bibitem[{{Gaidos} {et~al.}(2015){Gaidos}, {Mann}, \& {Ansdell}}]{Gaidos:2015}
{Gaidos}, E., {Mann}, A.~W., \& {Ansdell}, M. 2015, ArXiv e-prints,
  arXiv:1511.06471

\bibitem[{{Hayden} {et~al.}(2015){Hayden}, {Bovy}, {Holtzman}, {Nidever},
  {Bird}, {Weinberg}, {Andrews}, {Allende Prieto}, {Anders}, {Beers},
  {Bizyaev}, {Chiappini}, {Cunha}, {Frinchaboy}, {Garc{\'{\i}}a-Her{\'n}andez},
  {Garc{\'{\i}}a P{\'e}rez}, {Girardi}, {Harding}, {Hearty}, {Johnson},
  {Majewski}, {M{\'e}sz{\'a}ros}, {Minchev}, {O'Connell}, {Pan}, {Robin},
  {Schiavon}, {Schneider}, {Schultheis}, {Shetrone}, {Skrutskie}, {Steinmetz},
  {Smith}, {Zamora}, \& {Zasowski}}]{Hayden:2015}
{Hayden}, M.~R., {Bovy}, J., {Holtzman}, J.~A., {et~al.} 2015, ArXiv e-prints

\bibitem[{{Huber} {et~al.}(2014){Huber}, {Silva Aguirre}, {Matthews},
  {Pinsonneault}, {Gaidos}, {Garc{\'{\i}}a}, {Hekker}, {Mathur}, {Mosser},
  {Torres}, {Bastien}, {Basu}, {Bedding}, {Chaplin}, {Demory}, {Fleming},
  {Guo}, {Mann}, {Rowe}, {Serenelli}, {Smith}, \& {Stello}}]{Huber:2014}
{Huber}, D., {Silva Aguirre}, V., {Matthews}, J.~M., {et~al.} 2014, \apjs, 211,
  2

\bibitem[{{Jontof-Hutter} {et~al.}(2015){Jontof-Hutter}, {Ford}, {Rowe},
  {Lissauer}, {Fabrycky}, {Van Laerhoven}, {Agol}, {Deck}, {Holczer}, \&
  {Mazeh}}]{Jontof:2015}
{Jontof-Hutter}, D., {Ford}, E.~B., {Rowe}, J.~F., {et~al.} 2015, ArXiv
  e-prints, arXiv:1512.02003

\bibitem[{{Kipping} {et~al.}(2014){Kipping}, {Torres}, {Buchhave}, {Kenyon},
  {Henze}, {Isaacson}, {Kolbl}, {Marcy}, {Bryson}, {Stassun}, \&
  {Bastien}}]{Kipping:2014}
{Kipping}, D.~M., {Torres}, G., {Buchhave}, L.~A., {et~al.} 2014, \apj, 795, 25

\bibitem[{{Kopparapu} {et~al.}(2013){Kopparapu}, {Ramirez}, {Kasting}, {Eymet},
  {Robinson}, {Mahadevan}, {Terrien}, {Domagal-Goldman}, {Meadows}, \&
  {Deshpande}}]{Kopparapu:2013}
{Kopparapu}, R.~K., {Ramirez}, R., {Kasting}, J.~F., {et~al.} 2013, \apj, 765,
  131

\bibitem[{{Lissauer} {et~al.}(2012){Lissauer}, {Marcy}, {Rowe}, {Bryson},
  {Adams}, {Buchhave}, {Ciardi}, {Cochran}, {Fabrycky}, {Ford}, {Fressin},
  {Geary}, {Gilliland}, {Holman}, {Howell}, {Jenkins}, {Kinemuchi}, {Koch},
  {Morehead}, {Ragozzine}, {Seader}, {Tanenbaum}, {Torres}, \&
  {Twicken}}]{Lissauer:2012}
{Lissauer}, J.~J., {Marcy}, G.~W., {Rowe}, J.~F., {et~al.} 2012, \apj, 750, 112

\bibitem[{{Lissauer} {et~al.}(2014){Lissauer}, {Marcy}, {Bryson}, {Rowe},
  {Jontof-Hutter}, {Agol}, {Borucki}, {Carter}, {Ford}, {Gilliland}, {Kolbl},
  {Star}, {Steffen}, \& {Torres}}]{Lissauer:2014}
{Lissauer}, J.~J., {Marcy}, G.~W., {Bryson}, S.~T., {et~al.} 2014, \apj, 784,
  44

\bibitem[{{Marcy} {et~al.}(2014){Marcy}, {Isaacson}, {Howard}, {Rowe},
  {Jenkins}, {Bryson}, {Latham}, {Howell}, {Gautier}, {Batalha}, {Rogers},
  {Ciardi}, {Fischer}, {Gilliland}, {Kjeldsen}, {Christensen-Dalsgaard},
  {Huber}, {Chaplin}, {Basu}, {Buchhave}, {Quinn}, {Borucki}, {Koch}, {Hunter},
  {Caldwell}, {Van Cleve}, {Kolbl}, {Weiss}, {Petigura}, {Seager}, {Morton},
  {Johnson}, {Ballard}, {Burke}, {Cochran}, {Endl}, {MacQueen}, {Everett},
  {Lissauer}, {Ford}, {Torres}, {Fressin}, {Brown}, {Steffen}, {Charbonneau},
  {Basri}, {Sasselov}, {Winn}, {Sanchis-Ojeda}, {Christiansen}, {Adams},
  {Henze}, {Dupree}, {Fabrycky}, {Fortney}, {Tarter}, {Holman}, {Tenenbaum},
  {Shporer}, {Lucas}, {Welsh}, {Orosz}, {Bedding}, {Campante}, {Davies},
  {Elsworth}, {Handberg}, {Hekker}, {Karoff}, {Kawaler}, {Lund}, {Lundkvist},
  {Metcalfe}, {Miglio}, {Silva Aguirre}, {Stello}, {White}, {Boss}, {Devore},
  {Gould}, {Prsa}, {Agol}, {Barclay}, {Coughlin}, {Brugamyer}, {Mullally},
  {Quintana}, {Still}, {Thompson}, {Morrison}, {Twicken}, {D{\'e}sert},
  {Carter}, {Crepp}, {H{\'e}brard}, {Santerne}, {Moutou}, {Sobeck}, {Hudgins},
  {Haas}, {Robertson}, {Lillo-Box}, \& {Barrado}}]{Marcy:2014}
{Marcy}, G.~W., {Isaacson}, H., {Howard}, A.~W., {et~al.} 2014, \apjs, 210, 20

\bibitem[{{Montet} {et~al.}(2015){Montet}, {Morton}, {Foreman-Mackey},
  {Johnson}, {Hogg}, {Bowler}, {Latham}, {Bieryla}, \& {Mann}}]{Montet:2015}
{Montet}, B.~T., {Morton}, T.~D., {Foreman-Mackey}, D., {et~al.} 2015, ArXiv
  e-prints

\bibitem[{{Morton}(2014)}]{Morton:2014b}
{Morton}, T. 2014, PhD thesis, California Institute of Technology

\bibitem[{{Morton}(2012)}]{Morton:2012}
{Morton}, T.~D. 2012, \apj, 761, 6

\bibitem[{{Morton}(2015{\natexlab{a}})}]{isochrones}
---. 2015{\natexlab{a}}, {isochrones: Stellar model grid package}, Astrophysics
  Source Code Library, ascl:1503.010

\bibitem[{{Morton}(2015{\natexlab{b}})}]{vespa}
---. 2015{\natexlab{b}}, {VESPA: False positive probabilities calculator},
  Astrophysics Source Code Library, ascl:1503.011

\bibitem[{{Morton} \& {Johnson}(2011)}]{Morton:2011b}
{Morton}, T.~D., \& {Johnson}, J.~A. 2011, \apj, 738, 170

\bibitem[{{Morton} \& {Swift}(2014)}]{MortonSwift:2014}
{Morton}, T.~D., \& {Swift}, J. 2014, \apj, 791, 10

\bibitem[{{Morton} \& {Winn}(2014)}]{MortonWinn:2014}
{Morton}, T.~D., \& {Winn}, J.~N. 2014, \apj, 796, 47

\bibitem[{{Moutou} {et~al.}(2014){Moutou}, {Almenara}, {D{\'{\i}}az}, {Alonso},
  {Deleuil}, {Guenther}, {Pasternacki}, {Aigrain}, {Baglin}, {Barge}, {Bonomo},
  {Bord{\'e}}, {Bouchy}, {Cabrera}, {Carpano}, {Cochran}, {Csizmadia}, {Deeg},
  {Dvorak}, {Endl}, {Erikson}, {Ferraz-Mello}, {Fridlund}, {Gandolfi},
  {Guillot}, {Hatzes}, {H{\'e}brard}, {Lovis}, {Lammer}, {MacQueen}, {Mazeh},
  {Ofir}, {Ollivier}, {P{\"a}tzold}, {Rauer}, {Rouan}, {Santerne}, {Schneider},
  {Tingley}, \& {Wuchterl}}]{Moutou:2014}
{Moutou}, C., {Almenara}, J.~M., {D{\'{\i}}az}, R.~F., {et~al.} 2014, \mnras,
  444, 2783

\bibitem[{{Muirhead} {et~al.}(2012){Muirhead}, {Johnson}, {Apps}, {Carter},
  {Morton}, {Fabrycky}, {Pineda}, {Bottom}, {Rojas-Ayala}, {Schlawin},
  {Hamren}, {Covey}, {Crepp}, {Stassun}, {Pepper}, {Hebb}, {Kirby}, {Howard},
  {Isaacson}, {Marcy}, {Levitan}, {Diaz-Santos}, {Armus}, \&
  {Lloyd}}]{Muirhead:2012}
{Muirhead}, P.~S., {Johnson}, J.~A., {Apps}, K., {et~al.} 2012, \apj, 747, 144

\bibitem[{{Petigura} {et~al.}(2013){Petigura}, {Howard}, \&
  {Marcy}}]{Petigura:2013}
{Petigura}, E.~A., {Howard}, A.~W., \& {Marcy}, G.~W. 2013, Proceedings of the
  National Academy of Science, 110, 19273

\bibitem[{{Pinsonneault} {et~al.}(2012){Pinsonneault}, {An},
  {Molenda-{\.Z}akowicz}, {Chaplin}, {Metcalfe}, \&
  {Bruntt}}]{Pinsonneault:2012}
{Pinsonneault}, M.~H., {An}, D., {Molenda-{\.Z}akowicz}, J., {et~al.} 2012,
  \apjs, 199, 30

\bibitem[{{Rowe} {et~al.}(2014){Rowe}, {Bryson}, {Marcy}, {Lissauer},
  {Jontof-Hutter}, {Mullally}, {Gilliland}, {Issacson}, {Ford}, {Howell},
  {Borucki}, {Haas}, {Huber}, {Steffen}, {Thompson}, {Quintana}, {Barclay},
  {Still}, {Fortney}, {Gautier}, {Hunter}, {Caldwell}, {Ciardi}, {Devore},
  {Cochran}, {Jenkins}, {Agol}, {Carter}, \& {Geary}}]{Rowe:2014}
{Rowe}, J.~F., {Bryson}, S.~T., {Marcy}, G.~W., {et~al.} 2014, \apj, 784, 45

\bibitem[{{Rowe} {et~al.}(2015){Rowe}, {Coughlin}, {Antoci}, {Barclay},
  {Batalha}, {Borucki}, {Burke}, {Bryson}, {Caldwell}, {Campbell},
  {Catanzarite}, {Christiansen}, {Cochran}, {Gilliland}, {Girouard}, {Haas},
  {He{\l}miniak}, {Henze}, {Hoffman}, {Howell}, {Huber}, {Hunter},
  {Jang-Condell}, {Jenkins}, {Klaus}, {Latham}, {Li}, {Lissauer}, {McCauliff},
  {Morris}, {Mullally}, {Ofir}, {Quarles}, {Quintana}, {Sabale}, {Seader},
  {Shporer}, {Smith}, {Steffen}, {Still}, {Tenenbaum}, {Thompson}, {Twicken},
  {Van Laerhoven}, {Wolfgang}, \& {Zamudio}}]{Rowe:2015}
{Rowe}, J.~F., {Coughlin}, J.~L., {Antoci}, V., {et~al.} 2015, \apjs, 217, 16

\bibitem[{{Santerne} {et~al.}(2012){Santerne}, {D{\'{\i}}az}, {Moutou},
  {Bouchy}, {H{\'e}brard}, {Almenara}, {Bonomo}, {Deleuil}, \&
  {Santos}}]{Santerne:2012}
{Santerne}, A., {D{\'{\i}}az}, R.~F., {Moutou}, C., {et~al.} 2012, \aap, 545,
  A76

\bibitem[{{Santerne} {et~al.}(2014){Santerne}, {H{\'e}brard}, {Deleuil},
  {Havel}, {Correia}, {Almenara}, {Alonso}, {Arnold}, {Barros}, {Behrend},
  {Bernasconi}, {Boisse}, {Bonomo}, {Bouchy}, {Bruno}, {Damiani},
  {D{\'{\i}}az}, {Gravallon}, {Guillot}, {Labrevoir}, {Montagnier}, {Moutou},
  {Rinner}, {Santos}, {Abe}, {Audejean}, {Bendjoya}, {Gillier}, {Gregorio},
  {Martinez}, {Michelet}, {Montaigut}, {Poncy}, {Rivet}, {Rousseau}, {Roy},
  {Suarez}, {Vanhuysse}, \& {Verilhac}}]{Santerne:2014}
{Santerne}, A., {H{\'e}brard}, G., {Deleuil}, M., {et~al.} 2014, \aap, 571, A37

\bibitem[{{Santerne} {et~al.}(2015){Santerne}, {Moutou}, {Tsantaki}, {Bouchy},
  {H{\'e}brard}, {Adibekyan}, {Almenara}, {Amard}, {Barros}, {Boisse},
  {Bonomo}, {Bruno}, {Courcol}, {Deleuil}, {Demangeon}, {D{\'{\i}}az},
  {Guillot}, {Havel}, {Montagnier}, {Rajpurohit}, {Rey}, \&
  {Santos}}]{Santerne:2015}
{Santerne}, A., {Moutou}, C., {Tsantaki}, M., {et~al.} 2015, ArXiv e-prints,
  arXiv:1511.00643

\bibitem[{{Schlegel} {et~al.}(1998){Schlegel}, {Finkbeiner}, \&
  {Davis}}]{Schlegel:1998}
{Schlegel}, D.~J., {Finkbeiner}, D.~P., \& {Davis}, M. 1998, \apj, 500, 525

\bibitem[{{Sliski} \& {Kipping}(2014)}]{Sliski:2014}
{Sliski}, D.~H., \& {Kipping}, D.~M. 2014, \apj, 788, 148

\bibitem[{{Steffen} {et~al.}(2012){Steffen}, {Fabrycky}, {Ford}, {Carter},
  {D{\'e}sert}, {Fressin}, {Holman}, {Lissauer}, {Moorhead}, {Rowe},
  {Ragozzine}, {Welsh}, {Batalha}, {Borucki}, {Buchhave}, {Bryson}, {Caldwell},
  {Charbonneau}, {Ciardi}, {Cochran}, {Endl}, {Everett}, {Gautier},
  {Gilliland}, {Girouard}, {Jenkins}, {Horch}, {Howell}, {Isaacson}, {Klaus},
  {Koch}, {Latham}, {Li}, {Lucas}, {MacQueen}, {Marcy}, {McCauliff}, {Middour},
  {Morris}, {Mullally}, {Quinn}, {Quintana}, {Shporer}, {Still}, {Tenenbaum},
  {Thompson}, {Twicken}, \& {Van Cleve}}]{Steffen:2012}
{Steffen}, J.~H., {Fabrycky}, D.~C., {Ford}, E.~B., {et~al.} 2012, \mnras, 421,
  2342

\bibitem[{{Steffen} {et~al.}(2013){Steffen}, {Fabrycky}, {Agol}, {Ford},
  {Morehead}, {Cochran}, {Lissauer}, {Adams}, {Borucki}, {Bryson}, {Caldwell},
  {Dupree}, {Jenkins}, {Robertson}, {Rowe}, {Seader}, {Thompson}, \&
  {Twicken}}]{Steffen:2013}
{Steffen}, J.~H., {Fabrycky}, D.~C., {Agol}, E., {et~al.} 2013, \mnras, 428,
  1077

\bibitem[{{Swift} {et~al.}(2013){Swift}, {Johnson}, {Morton}, {Crepp},
  {Montet}, {Fabrycky}, \& {Muirhead}}]{Swift:2013}
{Swift}, J.~J., {Johnson}, J.~A., {Morton}, T.~D., {et~al.} 2013, \apj, 764,
  105

\bibitem[{{Torres} {et~al.}(2015){Torres}, {Kipping}, {Fressin}, {Caldwell},
  {Twicken}, {Ballard}, {Batalha}, {Bryson}, {Ciardi}, {Henze}, {Howell},
  {Isaacson}, {Jenkins}, {Muirhead}, {Newton}, {Petigura}, {Barclay},
  {Borucki}, {Crepp}, {Everett}, {Horch}, {Howard}, {Kolbl}, {Marcy},
  {McCauliff}, \& {Quintana}}]{Torres:2015}
{Torres}, G., {Kipping}, D.~M., {Fressin}, F., {et~al.} 2015, \apj, 800, 99

\end{thebibliography}
\clearpage

\begin{figure*}[p]
\begin{center}
\includegraphics[width=7in]{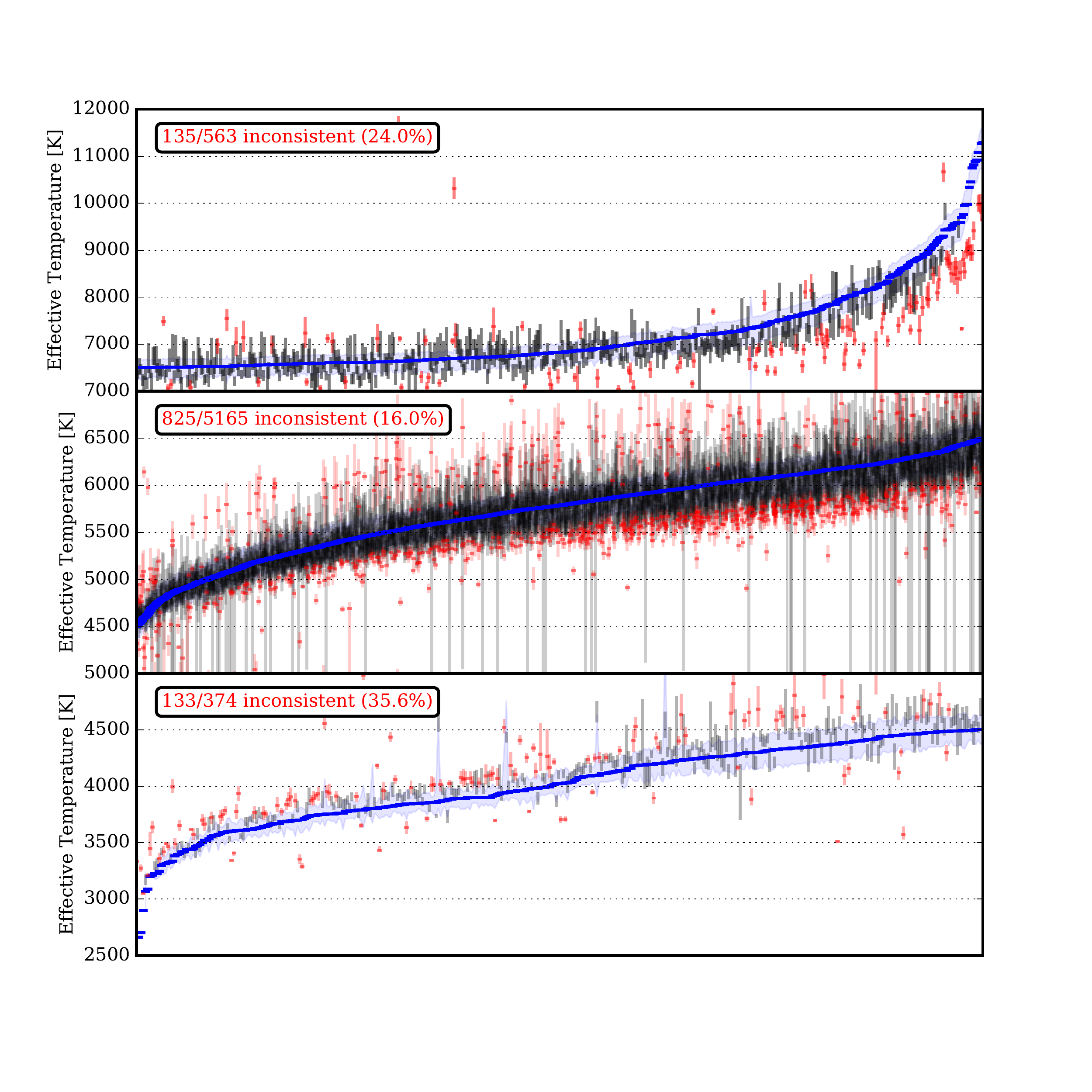}
\end{center}
\caption{Comparison between effective temperatures estimated from the
  \isochrones\ analysis in this work and those from the
  \kepler\ stellar parameters catalog \citep[][hereafter
    \citetalias{Huber:2014}]{Huber:2014}.  Bottom panel shows stars
  for which \citetalias{Huber:2014} predicts $T_{\rm eff} < 4500$\,K,
  middle spans $4500\,{\rm K} < T_{\rm eff} < 6500\,{\rm K}$, and top
  has $T_{\rm eff} > 6500$\,K. Blue horizontal bold lines are the
  \citetalias{Huber:2014} values in sorted order; blue shading
  represents the error bars from \citetalias{Huber:2014}.  Vertical
  lines span the 1$\sigma$ credible region of the \isochrones\ fits;
  these lines are grey if they overlap with the
  \citetalias{Huber:2014} 1$\sigma$ region and red (with the median
  marked by a point) if they are inconsistent.  This comparison shows
  that the stellar parameters estimated in this work are broadly
  consistent with \citetalias{Huber:2014}, though less so for the
  coolest and hottest stars.
\figlabel{starsteff}}
\end{figure*}

\begin{figure*}[p]
\begin{center}
\includegraphics[width=7in]{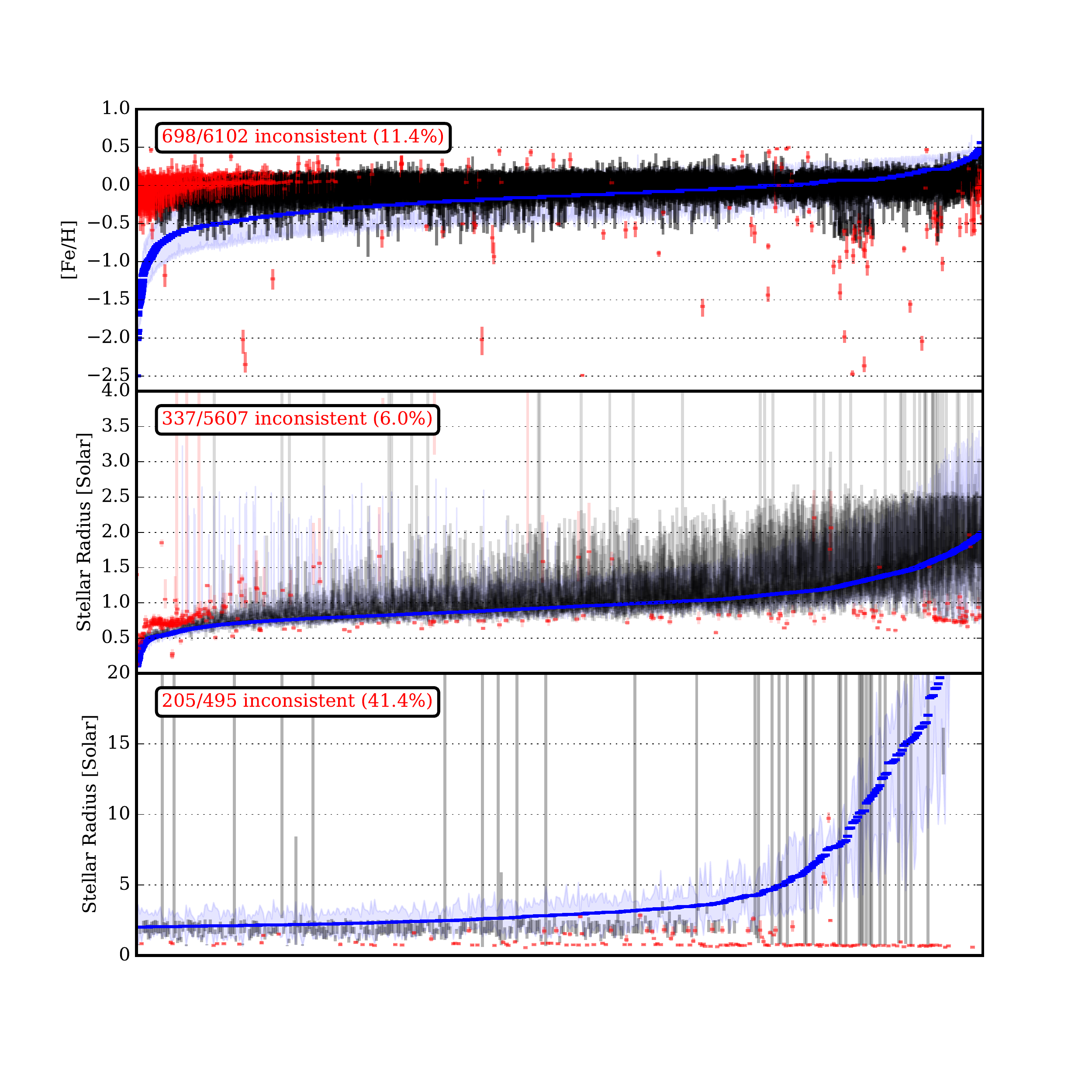}
\end{center}
\caption{Comparison between metallicities and radii estimated from the
  \isochrones\ analysis in this work and those from the
  \kepler\ stellar parameters catalog \citep[][hereafter
    \citetalias{Huber:2014}]{Huber:2014}.  Top panel shows metallicity
  for all stars in the sample.  The middle panel shows stars for which
  \citetalias{Huber:2014} estimates $R_\star < 2\,R_\odot$, and the
  bottom shows $R_\star > 2\,R_\odot$. Blue horizontal bold lines are
  the \citetalias{Huber:2014} values in sorted order; blue shading
  represents the error bars from \citetalias{Huber:2014}.  Vertical
  lines span the 1$\sigma$ credible region of the \isochrones\ fits;
  these lines are grey if they overlap with the
  \citetalias{Huber:2014} 1$\sigma$ region and red (with the median
  marked by a point) if they are inconsistent.  This comparison shows
  that the stellar parameters estimated in this work are broadly
  consistent with \citetalias{Huber:2014}, though less so for the more
  evolved stars.  The metallicity estimates of the
  \isochrones\ calculations are driven by the use of the local
  metallicity prior \citep[][\tab{priors}]{Hayden:2015,
    Casagrande:2011}.
\figlabel{starsfehradius}}
\end{figure*}

\begin{figure*}[p]
\begin{center}
\includegraphics[width=7in]{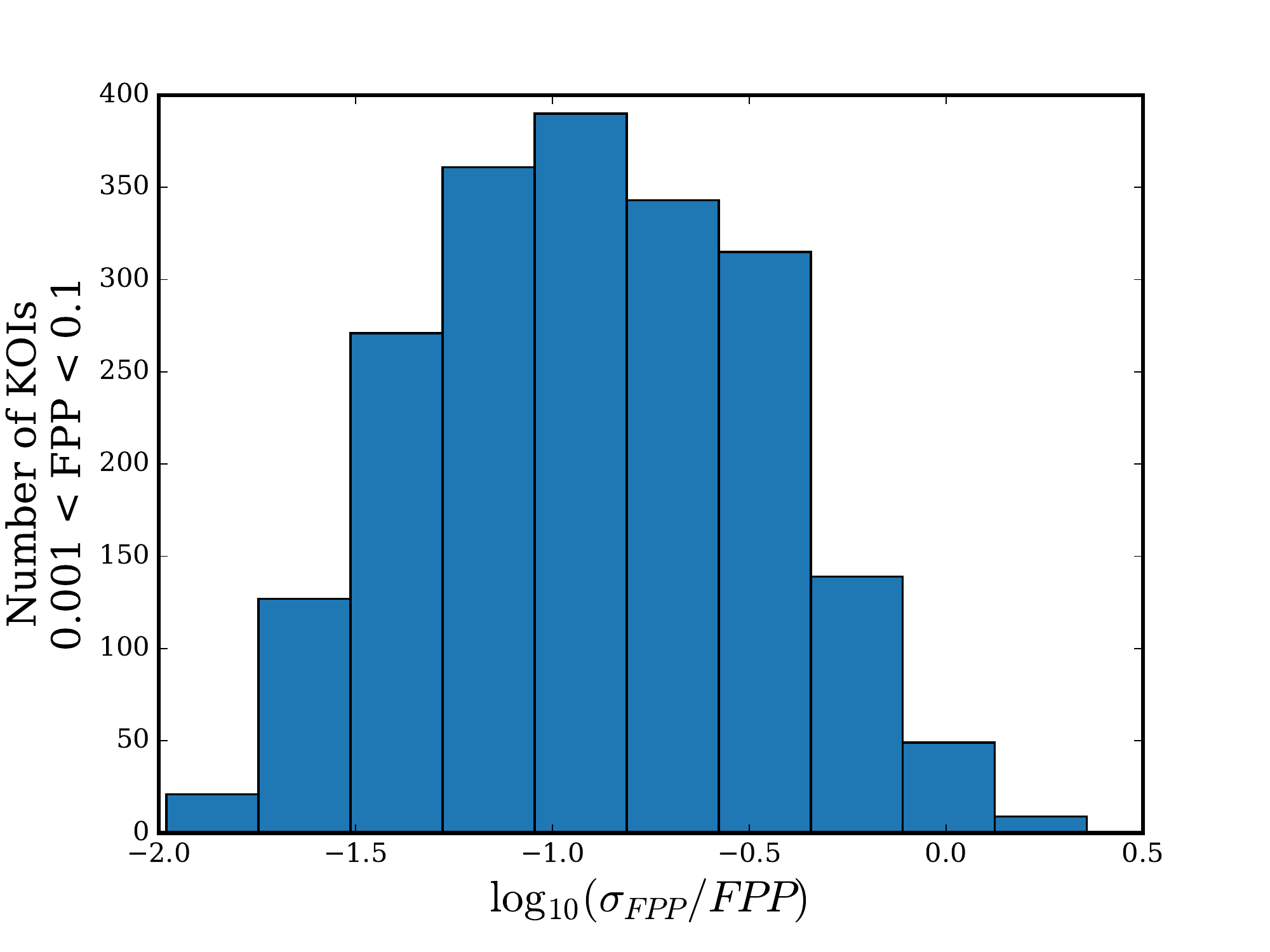}
\end{center}
\caption{Fractional uncertainties for KOIs with FPPs between
  0.001 and 0.1; that is, within an order of magnitude of the 
  validation threshold.  FPP values and uncertainties 
  are determined by the mean and standard deviations 
  of \vespa\ calculations based on 
  10 bootstrap resamplings (see \sect{methods:fpp}) of 
  a single set of simulated populations for each KOI.
  \figlabel{sigmafpp}}
\end{figure*}

\begin{figure*}[p]
\begin{center}
\includegraphics[width=7in]{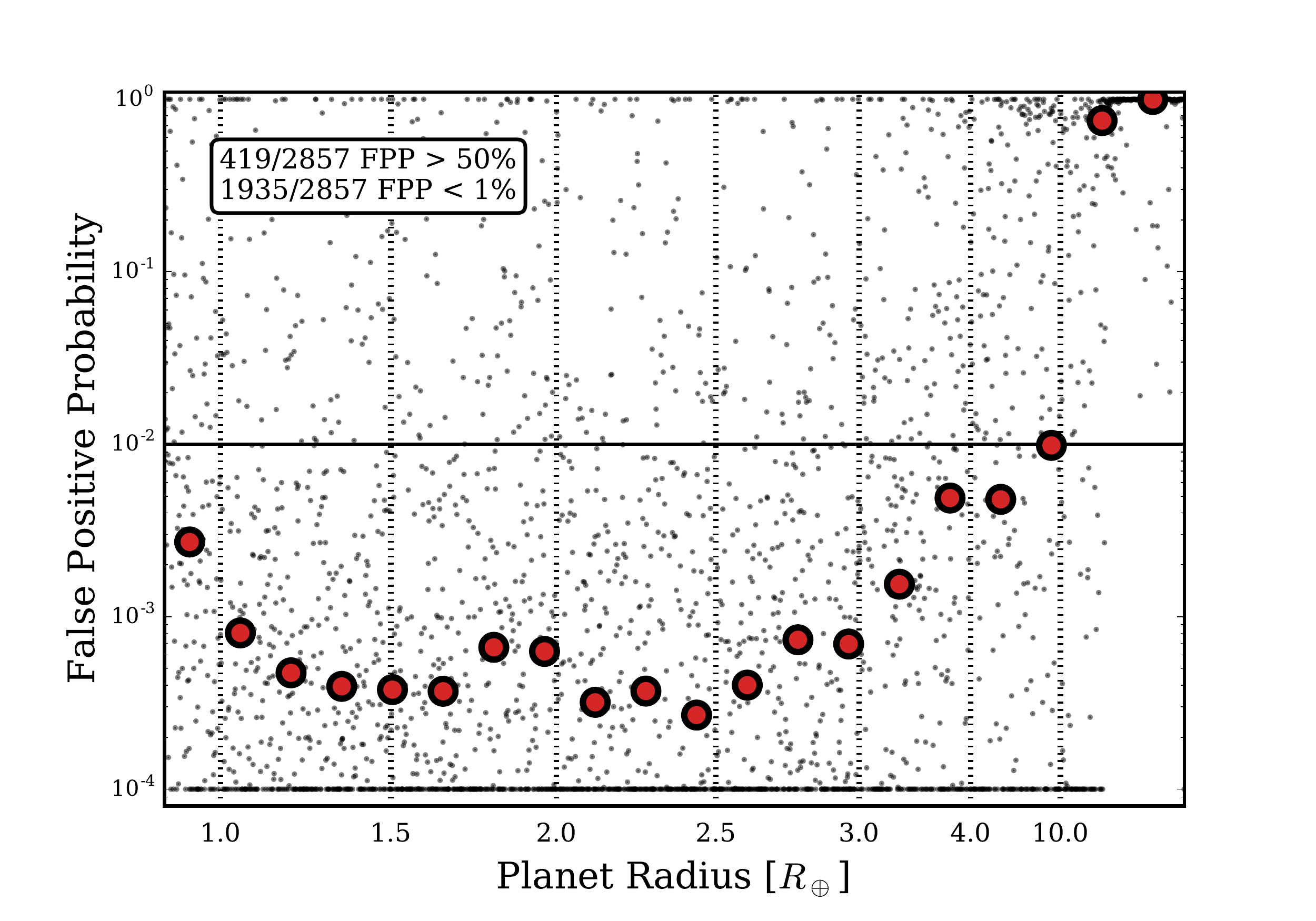}
\end{center}
\caption{False positive probabilities of all CANDIDATE or CONFIRMED
  KOIs for which we consider the \vespa\ calculations to be reliable
  (floored at $10^{-4}$ for visualization purposes), meaning they are
  considered to be reliably located on the target star ($ {\rm Pr} >
  0.99$, with ``score'' $> \posprobthresh$) according to pixel-level analysis
  \citep{Bryson:KSCI}, and have MES $>10$.  Of these \nreliable\ KOIs,
  \nval\ have FPPs less than 1\% (\nvalnew\ of which have not yet been
  dispositioned as CONFIRMED).  Noteably, \nreliableFP\ are likely
  false positives (FPP $> 0.5$), consistent with the
  \citet{Morton:2011b} and \citet{Fressin:2013} \emph{a priori}
  estimates of the overall \kepler\ candidate false positive rate.
  Red circles correspond to median FPP values in equal-sized bins. 
  \figlabel{fppall}}
\end{figure*}

\begin{figure*}[p]
\begin{center}
\includegraphics[width=7in]{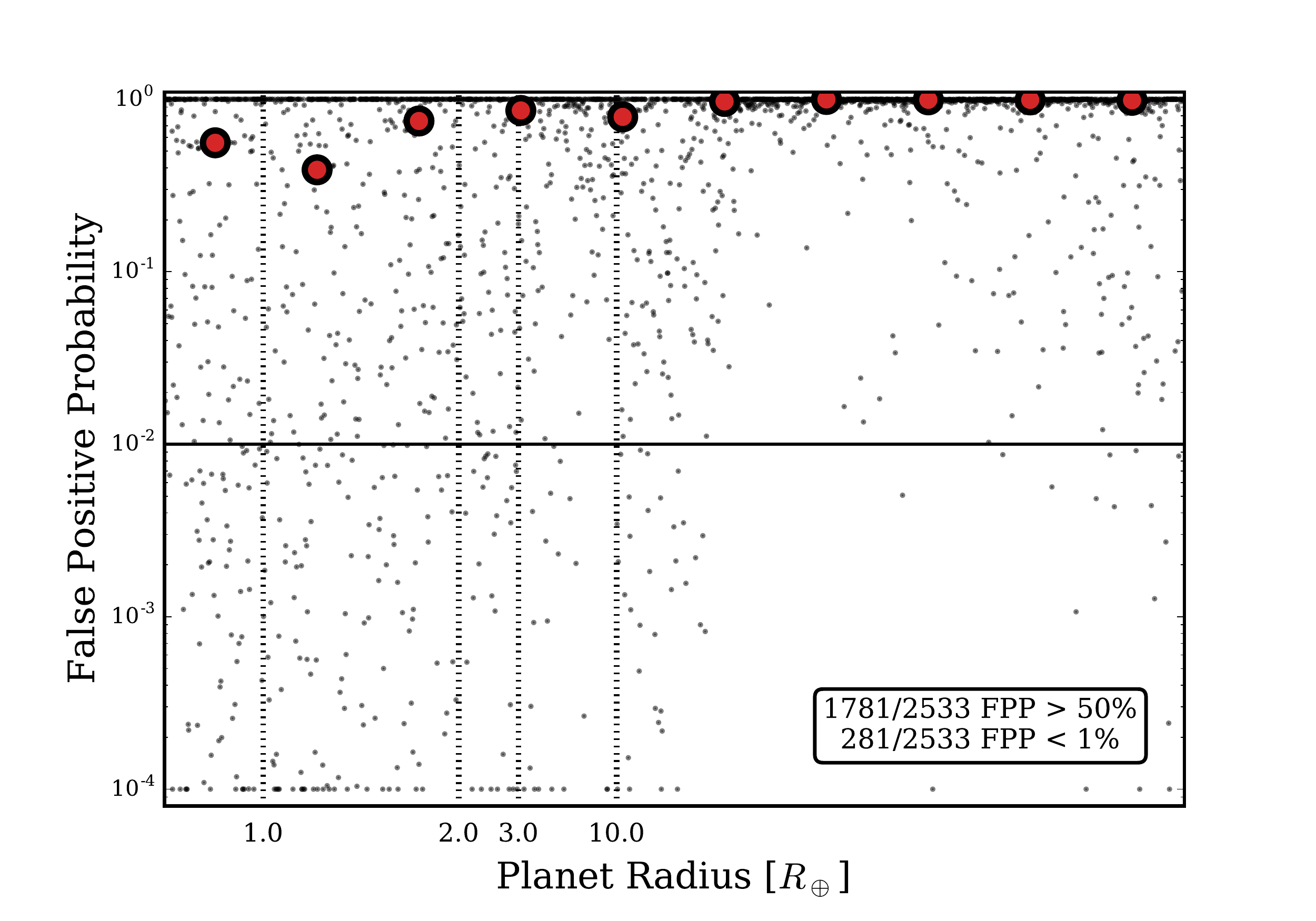}
\end{center}
\caption{Same as \Fig{fppall}, but for KOIs currently dispositioned
  as FALSE POSITIVE. The vast majority are also identified by \vespa\ as likely
  false positives.  There are also some that have low FPPs, but this can be
  explained by the fact that many of the reasons for dispositioning a 
  KOI as a FALSE POSITIVE also invalidate assumptions made by \vespa; 
  for example, that the signal is spatially coincident with the target 
  star (see \Sect{results}). This figure illustrates that \vespa\ is
  effective (though not 100\% efficient) at recovering known false positives.  
  \figlabel{fppfps}}
\end{figure*}
 
\begin{figure*}[p]
\begin{center}
\includegraphics[width=7in]{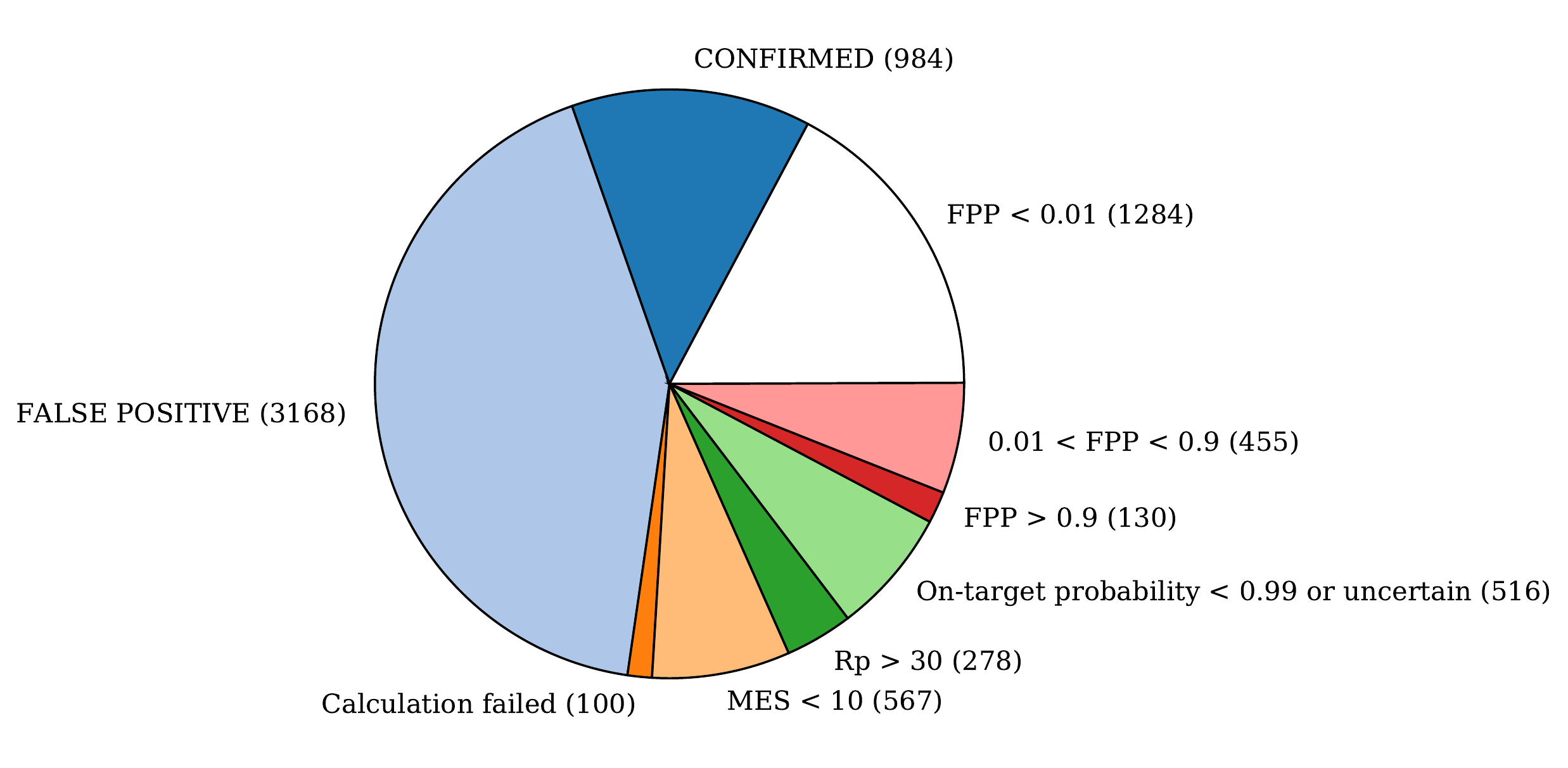}
\end{center}
\caption{A summary of how the the calculations presented in this paper
  advance our understanding of the true nature of KOIs.  More than
  half of all KOIs to date have already been dispositioned FALSE
  POSITIVE or CONFIRMED.  Those dispositioned CANDIDATE we further
  categorize according to their reliability.  The ``no reliable
  calculation'' category means that the \vespa\ calculation was
  not successful.  ``On-target
  probability uncertain'' indicates that the positional probability
  calculations of \citet{Bryson:KSCI} are not reliable (score $<
  \posprobthresh$).  ``On-target probability $<$ 0.99'' means that the positional
  probability calculations indicate that there is a non-negligible
  chance that the source of the transit signal is not at the position
  of the KOI.  The remaining three categories are all CANDIDATE KOIs
  reliably confirmed to be located at the presumed target star,
  grouped by false positive probability.  \nvalnew\ of these are new
  planet validations.
  \figlabel{fpppie}}
\end{figure*}

\begin{figure*}[p]
\begin{center}
\includegraphics[width=7in]{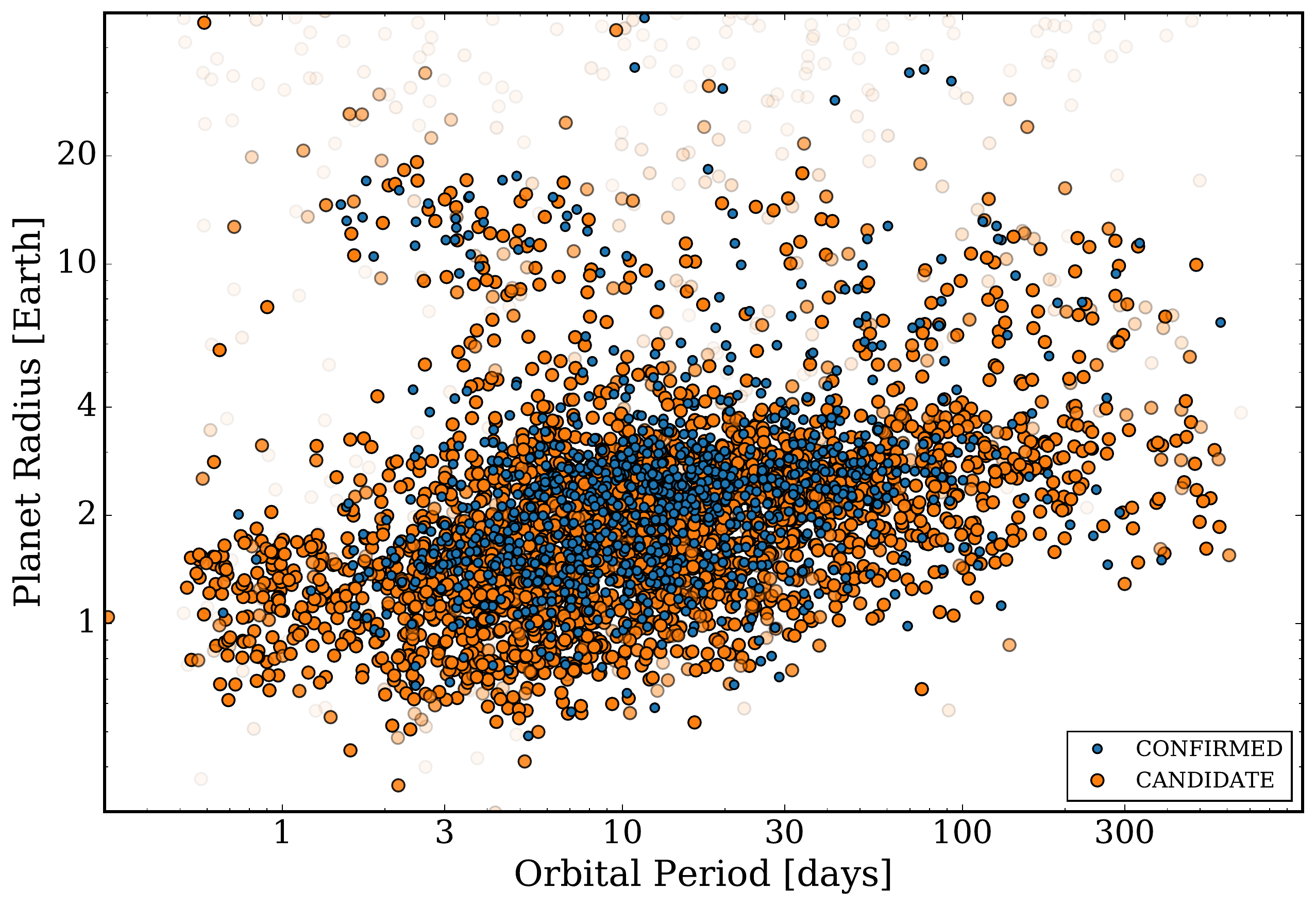}
\end{center}
\caption{Periods and radii of KOIs with CANDIDATE and CONFIRMED disposition.  
  Blue circles have previously been identified as CONFIRMED.  Candidates are orange circles, 
  shaded by false positive probability, with 
  a transparent circle representing a high FPP. 
  \figlabel{rpcand}}
\end{figure*}

\begin{figure*}[p]
\begin{center}
\includegraphics[width=7in]{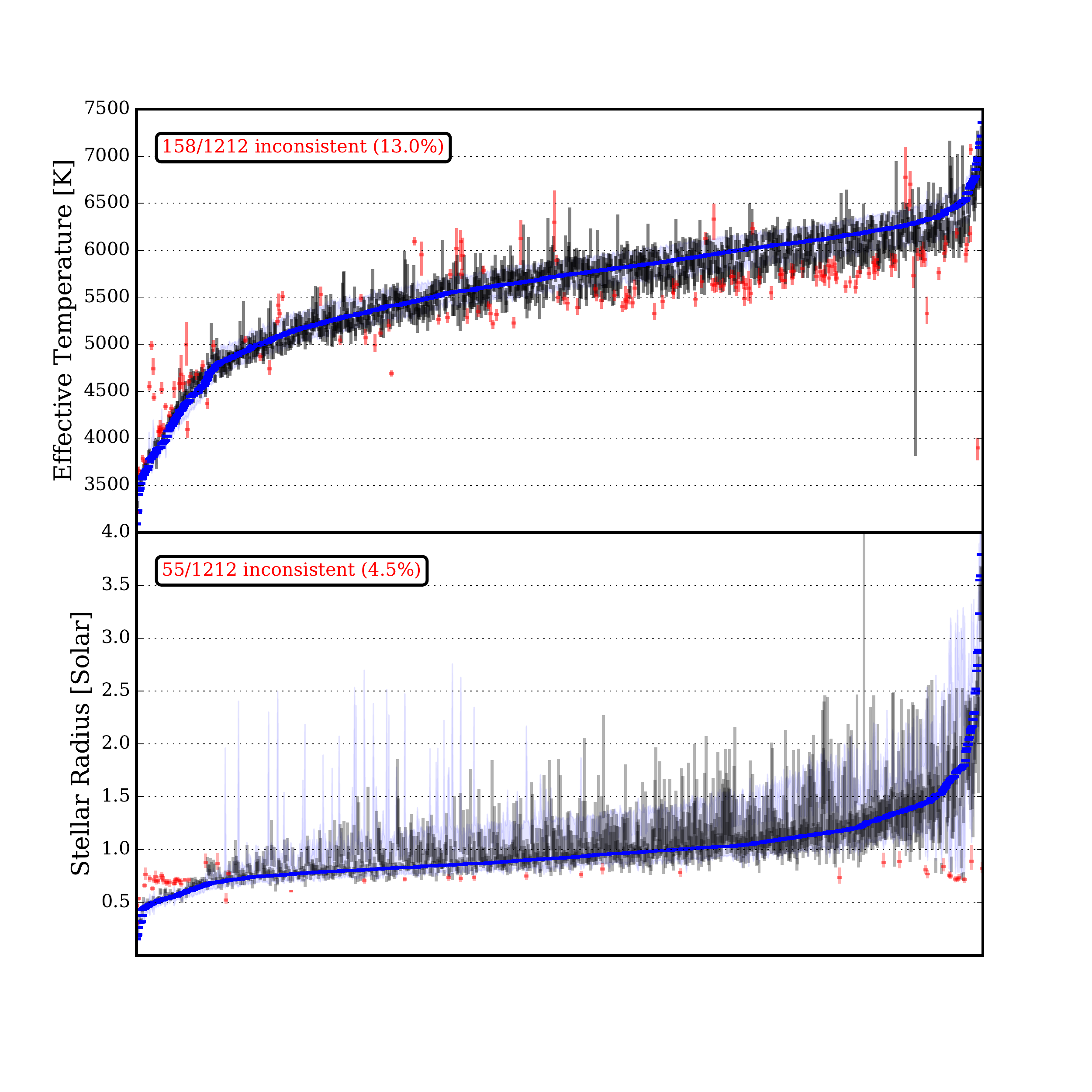}
\end{center}
\caption{Comparison between temperatures and radii estimated from the
  \isochrones\ analysis in this work and those from the
  \kepler\ stellar parameters catalog \citepalias{Huber:2014}, for the 
  sample of stars hosting planets validated in this study.  As 
  seen in previous figures, the \isochrones\ analysis tends to 
  over-predict the stellar radii compared to \citetalias{Huber:2014} 
  for the coolest stars.  A small number of stars are also estimated by 
  the photometric analysis of \citetalias{Huber:2014} to be evolved, 
  but not by \isochrones.
  \figlabel{validatedcompare}}
\end{figure*}


\begin{deluxetable*}{lccccccccccc}
\tablewidth{0pt}
\tabletypesize{\scriptsize}
\tablecaption{Stellar Properties
\tablabel{stars}}
\tablehead{\colhead{KOI} &
    \colhead{$M_\star$} &
    \colhead{$R_\star$} &
    \colhead{$T_{\rm eff}$} &
    \colhead{$\log g$} &
    \colhead{[Fe/H]} &
    \colhead{Age} &
    \colhead{$d$} &
    \colhead{$A_V$}&
    \colhead{$\pi\left(T_{\rm eff}\right)$} &
    \colhead{$\pi\left(\log g\right)$} &
    \colhead{$\pi\left({\rm [Fe/H]}\right)$} \\
    \colhead{} &
    \colhead{($M_\odot$)} &
    \colhead{($R_\odot$)} &
    \colhead{(K)} &
    \colhead{(cgs)} &
    \colhead{(dex)} &
    \colhead{(Gyr)} &
    \colhead{(pc)} &
    \colhead{(mag)} &
    \colhead{} &
    \colhead{} &
    \colhead{} 
    }
\startdata
K00757.03 &${ 0.80 }^{ +0.04 }_{ -0.04 }$&${ 0.76 }^{ +0.03 }_{ -0.04 }$&${ 5017 }^{ +54 }_{ -57 }$&${ 4.59 }^{ +0.03 }_{ -0.05 }$&${ -0.04 }^{ +0.15 }_{ -0.17 }$&${ 9.69 }^{ +0.37 }_{ -0.41 }$&${ 879 }^{ +42 }_{ -49 }$&${ 0.10 }^{ +0.06 }_{ -0.06 }$&--&--&--\\ 
K00758.01 &${ 0.79 }^{ +0.04 }_{ -0.04 }$&${ 0.74 }^{ +0.03 }_{ -0.04 }$&${ 4913 }^{ +94 }_{ -87 }$&${ 4.60 }^{ +0.02 }_{ -0.04 }$&${ -0.03 }^{ +0.15 }_{ -0.17 }$&${ 9.67 }^{ +0.36 }_{ -0.38 }$&${ 651 }^{ +32 }_{ -37 }$&${ 0.17 }^{ +0.12 }_{ -0.11 }$&--&--&--\\ 
K00759.02 &${ 0.84 }^{ +0.05 }_{ -0.05 }$&${ 0.81 }^{ +0.06 }_{ -0.06 }$&${ 5328 }^{ +60 }_{ -55 }$&${ 4.56 }^{ +0.04 }_{ -0.08 }$&${ -0.14 }^{ +0.16 }_{ -0.19 }$&${ 9.64 }^{ +0.39 }_{ -0.43 }$&${ 775 }^{ +64 }_{ -61 }$&${ 0.06 }^{ +0.05 }_{ -0.04 }$&--&--&--\\ 
K00760.01 &${ 1.06 }^{ +0.07 }_{ -0.06 }$&${ 1.07 }^{ +0.25 }_{ -0.10 }$&${ 5895 }^{ +86 }_{ -102 }$&${ 4.40 }^{ +0.08 }_{ -0.17 }$&${ 0.06 }^{ +0.15 }_{ -0.15 }$&${ 9.59 }^{ +0.20 }_{ -0.35 }$&${ 1356 }^{ +311 }_{ -141 }$&${ 0.14 }^{ +0.05 }_{ -0.08 }$&--&--&--\\ 
K00761.01 &${ 0.94 }^{ +0.04 }_{ -0.04 }$&${ 0.92 }^{ +0.19 }_{ -0.06 }$&${ 5505 }^{ +69 }_{ -90 }$&${ 4.49 }^{ +0.05 }_{ -0.17 }$&${ 0.09 }^{ +0.13 }_{ -0.13 }$&${ 9.71 }^{ +0.35 }_{ -0.46 }$&${ 1051 }^{ +223 }_{ -75 }$&${ 0.32 }^{ +0.04 }_{ -0.09 }$&--&--&--\\ 
K00762.01 &${ 1.03 }^{ +0.09 }_{ -0.08 }$&${ 1.05 }^{ +0.25 }_{ -0.12 }$&${ 5887 }^{ +165 }_{ -179 }$&${ 4.41 }^{ +0.09 }_{ -0.17 }$&${ 0.01 }^{ +0.14 }_{ -0.17 }$&${ 9.58 }^{ +0.24 }_{ -0.34 }$&${ 1330 }^{ +317 }_{ -168 }$&${ 0.28 }^{ +0.12 }_{ -0.15 }$&--&--&--\\ 
K00763.01 &${ 0.99 }^{ +0.04 }_{ -0.05 }$&${ 0.98 }^{ +0.18 }_{ -0.08 }$&${ 5710 }^{ +73 }_{ -91 }$&${ 4.45 }^{ +0.06 }_{ -0.15 }$&${ 0.05 }^{ +0.14 }_{ -0.17 }$&${ 9.67 }^{ +0.27 }_{ -0.39 }$&${ 1304 }^{ +241 }_{ -109 }$&${ 0.15 }^{ +0.04 }_{ -0.08 }$&--&--&--\\ 
K00764.01 &${ 0.91 }^{ +0.04 }_{ -0.04 }$&${ 0.87 }^{ +0.07 }_{ -0.05 }$&${ 5327 }^{ +53 }_{ -70 }$&${ 4.53 }^{ +0.04 }_{ -0.09 }$&${ 0.15 }^{ +0.15 }_{ -0.16 }$&${ 9.67 }^{ +0.37 }_{ -0.44 }$&${ 907 }^{ +76 }_{ -52 }$&${ 0.21 }^{ +0.03 }_{ -0.07 }$&--&--&--\\ 
K00765.01 &${ 0.87 }^{ +0.04 }_{ -0.05 }$&${ 0.82 }^{ +0.06 }_{ -0.05 }$&${ 5352 }^{ +61 }_{ -61 }$&${ 4.55 }^{ +0.03 }_{ -0.07 }$&${ -0.06 }^{ +0.15 }_{ -0.17 }$&${ 9.59 }^{ +0.40 }_{ -0.40 }$&${ 877 }^{ +70 }_{ -59 }$&${ 0.09 }^{ +0.05 }_{ -0.06 }$&--&--&--\\ 
K00766.01 &${ 1.15 }^{ +0.10 }_{ -0.07 }$&${ 1.20 }^{ +0.32 }_{ -0.13 }$&${ 6038 }^{ +98 }_{ -109 }$&${ 4.34 }^{ +0.08 }_{ -0.18 }$&${ 0.14 }^{ +0.15 }_{ -0.14 }$&${ 9.50 }^{ +0.15 }_{ -0.29 }$&${ 1778 }^{ +463 }_{ -208 }$&${ 0.14 }^{ +0.05 }_{ -0.08 }$&--&--&--\\ 
K00767.01 &${ 0.99 }^{ +0.05 }_{ -0.06 }$&${ 0.99 }^{ +0.22 }_{ -0.09 }$&${ 5709 }^{ +89 }_{ -134 }$&${ 4.44 }^{ +0.07 }_{ -0.17 }$&${ 0.07 }^{ +0.14 }_{ -0.15 }$&${ 9.68 }^{ +0.25 }_{ -0.41 }$&${ 980 }^{ +218 }_{ -98 }$&${ 0.36 }^{ +0.06 }_{ -0.12 }$&--&--&--\\ 
K00768.01 &${ 0.81 }^{ +0.03 }_{ -0.04 }$&${ 0.77 }^{ +0.03 }_{ -0.03 }$&${ 5035 }^{ +47 }_{ -49 }$&${ 4.58 }^{ +0.02 }_{ -0.05 }$&${ 0.00 }^{ +0.13 }_{ -0.15 }$&${ 9.68 }^{ +0.36 }_{ -0.40 }$&${ 857 }^{ +38 }_{ -39 }$&${ 0.08 }^{ +0.05 }_{ -0.06 }$&--&--&--\\ 
K00769.01 &${ 0.97 }^{ +0.05 }_{ -0.04 }$&${ 0.95 }^{ +0.17 }_{ -0.06 }$&${ 5624 }^{ +75 }_{ -91 }$&${ 4.47 }^{ +0.05 }_{ -0.15 }$&${ 0.08 }^{ +0.15 }_{ -0.14 }$&${ 9.65 }^{ +0.32 }_{ -0.43 }$&${ 1139 }^{ +200 }_{ -87 }$&${ 0.15 }^{ +0.05 }_{ -0.08 }$&--&--&--\\ 
K00770.01 &${ 0.93 }^{ +0.04 }_{ -0.05 }$&${ 0.89 }^{ +0.10 }_{ -0.06 }$&${ 5565 }^{ +88 }_{ -106 }$&${ 4.51 }^{ +0.04 }_{ -0.10 }$&${ -0.02 }^{ +0.15 }_{ -0.17 }$&${ 9.63 }^{ +0.33 }_{ -0.41 }$&${ 1114 }^{ +127 }_{ -77 }$&${ 0.16 }^{ +0.07 }_{ -0.10 }$&--&--&--\\ 
K00772.01 &${ 1.05 }^{ +0.06 }_{ -0.06 }$&${ 1.05 }^{ +0.21 }_{ -0.09 }$&${ 5917 }^{ +100 }_{ -85 }$&${ 4.41 }^{ +0.07 }_{ -0.14 }$&${ 0.01 }^{ +0.13 }_{ -0.17 }$&${ 9.56 }^{ +0.22 }_{ -0.32 }$&${ 1369 }^{ +274 }_{ -127 }$&${ 0.09 }^{ +0.07 }_{ -0.06 }$&--&--&--\\ 
K00773.01 &${ 0.95 }^{ +0.04 }_{ -0.05 }$&${ 0.92 }^{ +0.11 }_{ -0.06 }$&${ 5627 }^{ +75 }_{ -80 }$&${ 4.50 }^{ +0.05 }_{ -0.11 }$&${ -0.01 }^{ +0.15 }_{ -0.18 }$&${ 9.61 }^{ +0.33 }_{ -0.39 }$&${ 1011 }^{ +128 }_{ -74 }$&${ 0.12 }^{ +0.05 }_{ -0.07 }$&--&--&--\\ 
K00774.01 &${ 0.97 }^{ +0.06 }_{ -0.07 }$&${ 0.96 }^{ +0.16 }_{ -0.09 }$&${ 5760 }^{ +109 }_{ -104 }$&${ 4.46 }^{ +0.07 }_{ -0.13 }$&${ -0.07 }^{ +0.15 }_{ -0.21 }$&${ 9.65 }^{ +0.27 }_{ -0.40 }$&${ 1170 }^{ +189 }_{ -113 }$&${ 0.12 }^{ +0.08 }_{ -0.08 }$&--&--&--\\ 
K00775.01 &${ 0.66 }^{ +0.02 }_{ -0.03 }$&${ 0.63 }^{ +0.02 }_{ -0.02 }$&${ 4242 }^{ +42 }_{ -36 }$&${ 4.66 }^{ +0.01 }_{ -0.02 }$&${ -0.00 }^{ +0.12 }_{ -0.09 }$&${ 9.55 }^{ +0.36 }_{ -0.35 }$&${ 336 }^{ +12 }_{ -15 }$&${ 0.05 }^{ +0.06 }_{ -0.03 }$&(4117, 92)&(4.66, 0.03)&(0.04, 0.14)\\ 
K00775.02 &${ 0.66 }^{ +0.02 }_{ -0.03 }$&${ 0.63 }^{ +0.02 }_{ -0.02 }$&${ 4241 }^{ +42 }_{ -37 }$&${ 4.66 }^{ +0.01 }_{ -0.02 }$&${ -0.00 }^{ +0.11 }_{ -0.09 }$&${ 9.54 }^{ +0.37 }_{ -0.32 }$&${ 336 }^{ +12 }_{ -15 }$&${ 0.04 }^{ +0.06 }_{ -0.03 }$&(4117, 92)&(4.66, 0.03)&(0.04, 0.14)\\ 
K00775.03 &${ 0.66 }^{ +0.02 }_{ -0.03 }$&${ 0.63 }^{ +0.02 }_{ -0.02 }$&${ 4241 }^{ +40 }_{ -34 }$&${ 4.66 }^{ +0.01 }_{ -0.02 }$&${ -0.01 }^{ +0.11 }_{ -0.09 }$&${ 9.53 }^{ +0.37 }_{ -0.32 }$&${ 336 }^{ +11 }_{ -14 }$&${ 0.04 }^{ +0.06 }_{ -0.03 }$&(4117, 92)&(4.66, 0.03)&(0.04, 0.14)\\ 
K00776.01 &${ 0.88 }^{ +0.04 }_{ -0.05 }$&${ 0.84 }^{ +0.07 }_{ -0.05 }$&${ 5355 }^{ +60 }_{ -72 }$&${ 4.54 }^{ +0.04 }_{ -0.08 }$&${ 0.00 }^{ +0.15 }_{ -0.17 }$&${ 9.65 }^{ +0.38 }_{ -0.42 }$&${ 978 }^{ +80 }_{ -65 }$&${ 0.14 }^{ +0.04 }_{ -0.08 }$&--&--&--\\ 
K00777.01 &${ 0.82 }^{ +0.04 }_{ -0.05 }$&${ 0.79 }^{ +0.05 }_{ -0.05 }$&${ 5195 }^{ +56 }_{ -71 }$&${ 4.57 }^{ +0.04 }_{ -0.07 }$&${ -0.08 }^{ +0.16 }_{ -0.20 }$&${ 9.69 }^{ +0.36 }_{ -0.41 }$&${ 817 }^{ +57 }_{ -60 }$&${ 0.17 }^{ +0.04 }_{ -0.08 }$&--&--&--\\ 
K00778.01 &${ 0.60 }^{ +0.04 }_{ -0.03 }$&${ 0.57 }^{ +0.03 }_{ -0.02 }$&${ 4192 }^{ +41 }_{ -43 }$&${ 4.69 }^{ +0.01 }_{ -0.02 }$&${ -0.27 }^{ +0.14 }_{ -0.14 }$&${ 9.52 }^{ +0.42 }_{ -0.33 }$&${ 302 }^{ +19 }_{ -18 }$&${ 0.06 }^{ +0.05 }_{ -0.04 }$&(4128, 100)&(4.70, 0.02)&(-0.36, 0.16)\\ 
K00779.01 &${ 0.97 }^{ +0.04 }_{ -0.05 }$&${ 0.95 }^{ +0.15 }_{ -0.07 }$&${ 5652 }^{ +83 }_{ -102 }$&${ 4.47 }^{ +0.06 }_{ -0.14 }$&${ 0.03 }^{ +0.13 }_{ -0.16 }$&${ 9.67 }^{ +0.29 }_{ -0.42 }$&${ 1233 }^{ +200 }_{ -100 }$&${ 0.20 }^{ +0.05 }_{ -0.09 }$&--&--&--\\ 
K00780.01 &${ 0.79 }^{ +0.04 }_{ -0.05 }$&${ 0.75 }^{ +0.03 }_{ -0.04 }$&${ 4945 }^{ +70 }_{ -71 }$&${ 4.60 }^{ +0.03 }_{ -0.04 }$&${ -0.05 }^{ +0.15 }_{ -0.18 }$&${ 9.67 }^{ +0.36 }_{ -0.39 }$&${ 651 }^{ +32 }_{ -38 }$&${ 0.14 }^{ +0.09 }_{ -0.09 }$&--&--&--\\ 
K00780.02 &${ 0.79 }^{ +0.04 }_{ -0.04 }$&${ 0.75 }^{ +0.03 }_{ -0.04 }$&${ 4944 }^{ +74 }_{ -71 }$&${ 4.60 }^{ +0.02 }_{ -0.04 }$&${ -0.05 }^{ +0.15 }_{ -0.17 }$&${ 9.66 }^{ +0.37 }_{ -0.39 }$&${ 651 }^{ +30 }_{ -36 }$&${ 0.14 }^{ +0.09 }_{ -0.09 }$&--&--&--\\ 
K00781.01 &${ 0.51 }^{ +0.02 }_{ -0.03 }$&${ 0.49 }^{ +0.02 }_{ -0.03 }$&${ 3701 }^{ +39 }_{ -40 }$&${ 4.77 }^{ +0.02 }_{ -0.02 }$&${ -0.02 }^{ +0.09 }_{ -0.09 }$&${ 9.63 }^{ +0.38 }_{ -0.39 }$&${ 243 }^{ +15 }_{ -17 }$&${ 0.14 }^{ +0.08 }_{ -0.08 }$&(3648, 65)&(4.79, 0.06)&(0.00, 0.14)\\ 
K00782.01 &${ 1.00 }^{ +0.05 }_{ -0.05 }$&${ 0.99 }^{ +0.21 }_{ -0.08 }$&${ 5723 }^{ +69 }_{ -88 }$&${ 4.45 }^{ +0.06 }_{ -0.16 }$&${ 0.06 }^{ +0.14 }_{ -0.16 }$&${ 9.66 }^{ +0.26 }_{ -0.41 }$&${ 1180 }^{ +253 }_{ -99 }$&${ 0.17 }^{ +0.04 }_{ -0.07 }$&--&--&--\\ 
K00783.01 &${ 0.94 }^{ +0.04 }_{ -0.05 }$&${ 0.91 }^{ +0.12 }_{ -0.06 }$&${ 5520 }^{ +73 }_{ -108 }$&${ 4.50 }^{ +0.05 }_{ -0.12 }$&${ 0.07 }^{ +0.16 }_{ -0.15 }$&${ 9.65 }^{ +0.35 }_{ -0.43 }$&${ 875 }^{ +113 }_{ -61 }$&${ 0.27 }^{ +0.05 }_{ -0.10 }$&--&--&--\\ 
K00784.01 &${ 0.59 }^{ +0.03 }_{ -0.04 }$&${ 0.57 }^{ +0.03 }_{ -0.03 }$&${ 4147 }^{ +45 }_{ -45 }$&${ 4.70 }^{ +0.02 }_{ -0.02 }$&${ -0.26 }^{ +0.13 }_{ -0.14 }$&${ 9.54 }^{ +0.38 }_{ -0.33 }$&${ 325 }^{ +19 }_{ -21 }$&${ 0.04 }^{ +0.06 }_{ -0.03 }$&(4059, 93)&(4.70, 0.02)&(-0.26, 0.18)\\ 
K00784.02 &${ 0.59 }^{ +0.03 }_{ -0.04 }$&${ 0.57 }^{ +0.03 }_{ -0.03 }$&${ 4147 }^{ +47 }_{ -46 }$&${ 4.70 }^{ +0.02 }_{ -0.02 }$&${ -0.26 }^{ +0.12 }_{ -0.14 }$&${ 9.53 }^{ +0.39 }_{ -0.32 }$&${ 325 }^{ +20 }_{ -22 }$&${ 0.05 }^{ +0.06 }_{ -0.03 }$&(4059, 93)&(4.70, 0.02)&(-0.26, 0.18)\\ 
K00785.01 &${ 0.87 }^{ +0.05 }_{ -0.05 }$&${ 0.84 }^{ +0.07 }_{ -0.06 }$&${ 5403 }^{ +97 }_{ -106 }$&${ 4.54 }^{ +0.04 }_{ -0.09 }$&${ -0.09 }^{ +0.16 }_{ -0.19 }$&${ 9.65 }^{ +0.38 }_{ -0.45 }$&${ 968 }^{ +90 }_{ -76 }$&${ 0.17 }^{ +0.09 }_{ -0.11 }$&--&--&--\\ 
K00786.01 &${ 1.08 }^{ +0.10 }_{ -0.06 }$&${ 1.11 }^{ +0.38 }_{ -0.11 }$&${ 5890 }^{ +77 }_{ -99 }$&${ 4.38 }^{ +0.08 }_{ -0.22 }$&${ 0.13 }^{ +0.15 }_{ -0.15 }$&${ 9.57 }^{ +0.18 }_{ -0.34 }$&${ 1342 }^{ +446 }_{ -145 }$&${ 0.21 }^{ +0.04 }_{ -0.08 }$&--&--&--\\ 
K00787.01 &${ 0.97 }^{ +0.06 }_{ -0.05 }$&${ 0.95 }^{ +0.20 }_{ -0.07 }$&${ 5654 }^{ +118 }_{ -108 }$&${ 4.47 }^{ +0.06 }_{ -0.16 }$&${ 0.03 }^{ +0.13 }_{ -0.16 }$&${ 9.65 }^{ +0.28 }_{ -0.40 }$&${ 1155 }^{ +238 }_{ -99 }$&${ 0.15 }^{ +0.10 }_{ -0.10 }$&--&--&--\\ 
K00787.02 &${ 0.96 }^{ +0.05 }_{ -0.05 }$&${ 0.94 }^{ +0.15 }_{ -0.07 }$&${ 5647 }^{ +123 }_{ -112 }$&${ 4.48 }^{ +0.05 }_{ -0.13 }$&${ 0.02 }^{ +0.13 }_{ -0.17 }$&${ 9.61 }^{ +0.32 }_{ -0.39 }$&${ 1142 }^{ +181 }_{ -93 }$&${ 0.15 }^{ +0.10 }_{ -0.10 }$&--&--&--\\ 
K00788.01 &${ 0.80 }^{ +0.04 }_{ -0.05 }$&${ 0.76 }^{ +0.03 }_{ -0.04 }$&${ 5021 }^{ +64 }_{ -62 }$&${ 4.58 }^{ +0.03 }_{ -0.04 }$&${ -0.04 }^{ +0.15 }_{ -0.16 }$&${ 9.70 }^{ +0.33 }_{ -0.39 }$&${ 666 }^{ +33 }_{ -37 }$&${ 0.11 }^{ +0.07 }_{ -0.08 }$&--&--&--\\ 
K00790.01 &${ 0.84 }^{ +0.04 }_{ -0.05 }$&${ 0.80 }^{ +0.06 }_{ -0.05 }$&${ 5261 }^{ +93 }_{ -81 }$&${ 4.56 }^{ +0.04 }_{ -0.07 }$&${ -0.09 }^{ +0.15 }_{ -0.18 }$&${ 9.67 }^{ +0.37 }_{ -0.44 }$&${ 825 }^{ +60 }_{ -60 }$&${ 0.12 }^{ +0.09 }_{ -0.08 }$&--&--&--\\ 
K00790.02 &${ 0.84 }^{ +0.04 }_{ -0.05 }$&${ 0.80 }^{ +0.05 }_{ -0.05 }$&${ 5257 }^{ +93 }_{ -74 }$&${ 4.56 }^{ +0.04 }_{ -0.07 }$&${ -0.09 }^{ +0.15 }_{ -0.18 }$&${ 9.65 }^{ +0.38 }_{ -0.42 }$&${ 823 }^{ +59 }_{ -58 }$&${ 0.11 }^{ +0.10 }_{ -0.08 }$&--&--&--\\ 
K00791.01 &${ 0.92 }^{ +0.04 }_{ -0.05 }$&${ 0.89 }^{ +0.09 }_{ -0.06 }$&${ 5559 }^{ +85 }_{ -82 }$&${ 4.51 }^{ +0.04 }_{ -0.10 }$&${ -0.05 }^{ +0.14 }_{ -0.17 }$&${ 9.60 }^{ +0.36 }_{ -0.39 }$&${ 951 }^{ +101 }_{ -68 }$&${ 0.10 }^{ +0.07 }_{ -0.07 }$&--&--&--\\ 
K00792.01 &${ 1.02 }^{ +0.09 }_{ -0.07 }$&${ 1.04 }^{ +0.21 }_{ -0.11 }$&${ 5908 }^{ +172 }_{ -172 }$&${ 4.42 }^{ +0.08 }_{ -0.14 }$&${ -0.03 }^{ +0.14 }_{ -0.18 }$&${ 9.58 }^{ +0.21 }_{ -0.32 }$&${ 1190 }^{ +245 }_{ -137 }$&${ 0.22 }^{ +0.13 }_{ -0.14 }$&--&--&--\\ 
K00793.01 &${ 1.12 }^{ +0.11 }_{ -0.09 }$&${ 1.17 }^{ +0.37 }_{ -0.15 }$&${ 6033 }^{ +119 }_{ -187 }$&${ 4.35 }^{ +0.10 }_{ -0.20 }$&${ 0.09 }^{ +0.14 }_{ -0.14 }$&${ 9.53 }^{ +0.18 }_{ -0.29 }$&${ 1257 }^{ +396 }_{ -168 }$&${ 0.46 }^{ +0.07 }_{ -0.14 }$&--&--&--\\ 
K00794.01 &${ 0.96 }^{ +0.06 }_{ -0.06 }$&${ 0.95 }^{ +0.16 }_{ -0.08 }$&${ 5703 }^{ +141 }_{ -149 }$&${ 4.47 }^{ +0.06 }_{ -0.13 }$&${ -0.03 }^{ +0.14 }_{ -0.18 }$&${ 9.64 }^{ +0.29 }_{ -0.38 }$&${ 976 }^{ +167 }_{ -94 }$&${ 0.23 }^{ +0.11 }_{ -0.13 }$&--&--&--\\ 
K00795.01 &${ 0.89 }^{ +0.06 }_{ -0.06 }$&${ 0.86 }^{ +0.08 }_{ -0.07 }$&${ 5493 }^{ +124 }_{ -119 }$&${ 4.53 }^{ +0.05 }_{ -0.09 }$&${ -0.11 }^{ +0.16 }_{ -0.19 }$&${ 9.64 }^{ +0.36 }_{ -0.43 }$&${ 1066 }^{ +98 }_{ -93 }$&${ 0.18 }^{ +0.11 }_{ -0.11 }$&--&--&--\\ 
K00796.01 &${ 0.80 }^{ +0.05 }_{ -0.06 }$&${ 0.76 }^{ +0.05 }_{ -0.06 }$&${ 5132 }^{ +107 }_{ -87 }$&${ 4.58 }^{ +0.04 }_{ -0.05 }$&${ -0.15 }^{ +0.16 }_{ -0.21 }$&${ 9.69 }^{ +0.36 }_{ -0.40 }$&${ 879 }^{ +60 }_{ -76 }$&${ 0.15 }^{ +0.12 }_{ -0.10 }$&--&--&--\\ 
K00797.01 &${ 0.93 }^{ +0.05 }_{ -0.06 }$&${ 0.90 }^{ +0.10 }_{ -0.07 }$&${ 5613 }^{ +112 }_{ -134 }$&${ 4.51 }^{ +0.05 }_{ -0.10 }$&${ -0.08 }^{ +0.15 }_{ -0.19 }$&${ 9.61 }^{ +0.33 }_{ -0.39 }$&${ 1190 }^{ +137 }_{ -100 }$&${ 0.21 }^{ +0.09 }_{ -0.12 }$&--&--&--\\ 
K00798.01 &${ 0.89 }^{ +0.04 }_{ -0.04 }$&${ 0.85 }^{ +0.08 }_{ -0.05 }$&${ 5342 }^{ +68 }_{ -105 }$&${ 4.53 }^{ +0.04 }_{ -0.09 }$&${ 0.07 }^{ +0.15 }_{ -0.14 }$&${ 9.68 }^{ +0.37 }_{ -0.44 }$&${ 950 }^{ +88 }_{ -58 }$&${ 0.34 }^{ +0.05 }_{ -0.11 }$&--&--&--\\ 
K00800.01 &${ 1.11 }^{ +0.08 }_{ -0.07 }$&${ 1.15 }^{ +0.31 }_{ -0.12 }$&${ 6003 }^{ +99 }_{ -137 }$&${ 4.36 }^{ +0.08 }_{ -0.18 }$&${ 0.09 }^{ +0.15 }_{ -0.16 }$&${ 9.54 }^{ +0.16 }_{ -0.31 }$&${ 1651 }^{ +438 }_{ -192 }$&${ 0.23 }^{ +0.05 }_{ -0.11 }$&--&--&--\\ 
K00800.02 &${ 1.11 }^{ +0.09 }_{ -0.07 }$&${ 1.15 }^{ +0.29 }_{ -0.13 }$&${ 6005 }^{ +102 }_{ -134 }$&${ 4.37 }^{ +0.08 }_{ -0.17 }$&${ 0.09 }^{ +0.15 }_{ -0.15 }$&${ 9.52 }^{ +0.18 }_{ -0.30 }$&${ 1651 }^{ +404 }_{ -190 }$&${ 0.23 }^{ +0.06 }_{ -0.11 }$&--&--&--\\ 
K00801.01 &${ 1.11 }^{ +0.11 }_{ -0.07 }$&${ 1.16 }^{ +0.33 }_{ -0.13 }$&${ 5963 }^{ +94 }_{ -126 }$&${ 4.36 }^{ +0.09 }_{ -0.19 }$&${ 0.15 }^{ +0.14 }_{ -0.15 }$&${ 9.53 }^{ +0.17 }_{ -0.32 }$&${ 1145 }^{ +327 }_{ -137 }$&${ 0.53 }^{ +0.05 }_{ -0.10 }$&--&--&--\\ 
K00802.01 &${ 0.96 }^{ +0.09 }_{ -0.07 }$&${ 0.95 }^{ +0.17 }_{ -0.09 }$&${ 5731 }^{ +255 }_{ -188 }$&${ 4.47 }^{ +0.07 }_{ -0.12 }$&${ -0.06 }^{ +0.15 }_{ -0.19 }$&${ 9.57 }^{ +0.30 }_{ -0.35 }$&${ 1247 }^{ +237 }_{ -130 }$&${ 0.24 }^{ +0.21 }_{ -0.16 }$&--&--&--
\enddata
\tablecomments{A portion of this table is shown for form and content.  
                The full table will be available online.}
\end{deluxetable*}


\begin{turnpage}

\begin{deluxetable*}{ccccccccccccccccccccccc}
\tablewidth{0pt}
\tabletypesize{\scriptsize}
\tablecaption{False Positive Probability Results
\tablabel{fpp}}
\tablehead{\colhead{KOI} &
    \colhead{$P$} &
    \colhead{TTV?} &
    \colhead{$R_p$} &
    \colhead{SNR} &
    \colhead{$\delta_{\rm sec}$\tablenotemark{a}} &
    \colhead{$r_{\rm excl}$\tablenotemark{b}} &
    \colhead{Pr$_{\rm EB}$\tablenotemark{c}} &
    \colhead{Pr$_{\rm EB2}$\tablenotemark{c}} &
    \colhead{Pr$_{\rm HEB}$\tablenotemark{c}} &
    \colhead{Pr$_{\rm HEB2}$\tablenotemark{c}} &
    \colhead{Pr$_{\rm BEB}$\tablenotemark{c}} &
    \colhead{Pr$_{\rm BEB2}$\tablenotemark{c}} &
    \colhead{Pr$_{\rm boxy}$\tablenotemark{d}} &
    \colhead{Pr$_{\rm long}$\tablenotemark{d}} &
    \colhead{$f_p$\tablenotemark{e}} &
    \colhead{$p_{\rm pos}$\tablenotemark{f}} &
    \colhead{$s_{\rm pos}$\tablenotemark{g}} &
    \colhead{Disp.\tablenotemark{h}} &
    \colhead{FPP\tablenotemark{i}} &
    \colhead{$\sigma_{\rm FPP}$\tablenotemark{j}} &
    \colhead{Failure\tablenotemark{k}} &
    \colhead{Kep\tablenotemark{l}} \\
    \colhead{} &
    \colhead{(d)} &
    \colhead{} &
    \colhead{($R_\oplus$)} &
    \colhead{} &
    \colhead{(ppm)} &
    \colhead{(\arcsec)} &
    \colhead{} &
    \colhead{} &
    \colhead{} &
    \colhead{} &
    \colhead{} &
    \colhead{} &
    \colhead{} &
    \colhead{} &
    \colhead{} &
    \colhead{} &
    \colhead{} &
    \colhead{} &
    \colhead{} &
    \colhead{} &
    \colhead{} &
    \colhead{}
    }
\startdata
K02360.01 & 2.304 & N & 6.64 & 22.6 & 42 & 0.50 & 0 & 0 & 0 & 0 & 0 & 0.0053 & 0 & 0.99 & 0.052 & 0.00 & 0.12 & FP & 1 & 0 & -- & -- \\ 
K02361.01 & 5.784 & N & 2.49 & 16.7 & 156 & 2.19 & 0.025 & 0.0016 & 0.0016 & 0.00026 & 0.0095 & 0.0054 & 0 & 0 & 0.201 & 1.00 & 0.25 & CA & 0.043 & 0.0026 & -- & -- \\ 
K02362.01 & 2.237 & N & 1.90 & 22.3 & 152 & 0.63 & 0 & 1.5e-05 & 0 & 0 & 0.0027 & 0.013 & 0 & 0 & 0.189 & 0.72 & 1.00 & FP & 0.016 & 0.00047 & -- & -- \\ 
K02362.02 & 11.085 & N & 2.32 & 15.8 & 316 & 0.87 & 0.0018 & 0.00013 & 8.8e-05 & 1.6e-05 & 0.00091 & 0.00018 & 0 & 0 & 0.215 & 1.00 & 0.64 & CA & 0.0031 & 0.00054 & -- & 1208b \\ 
K02363.01 & 3.139 & N & 1.03 & 19.6 & 51 & 0.90 & 0 & 0.00025 & 0 & 2.7e-05 & 0.029 & 0.015 & 0 & 0 & 0.101 & 1.00 & 0.88 & CA & 0.044 & 0.0016 & -- & -- \\ 
K02364.01 & 5.242 & N & 1.50 & 18.1 & 73 & 0.54 & 0 & 0 & 0 & 0 & 0 & 0 & 0.067 & 0.12 & 0.146 & 1.00 & 1.00 & CA & 0.19 & 0.057 & -- & -- \\ 
K02365.01 & 35.968 & N & 1.98 & 19.4 & 81 & 1.44 & 0.02 & 0.0004 & 0.00072 & 4.1e-05 & 0.0023 & 0.0012 & 1.2e-06 & 0 & 0.208 & 1.00 & 0.88 & PL & 0.024 & 0.0021 & -- & -- \\ 
K02365.02 & 110.974 & N & 1.45 & 10.6 & 65 & 3.30 & 0 & 0 & 0 & 0 & 0.00031 & 4.1e-05 & 0.001 & 0 & 0.145 & 1.00 & 0.59 & PL & 0.0013 & 0.0001 & -- & -- \\ 
K02366.01 & 25.369 & N & 1.51 & 18.4 & 32 & 1.32 & 0.0016 & 0 & 4.1e-06 & 0 & 0 & 0 & 0.005 & 0 & 0.152 & 1.00 & 1.00 & CA & 0.0066 & 0.00055 & -- & 1209b \\ 
K02367.01 & 6.892 & N & 1.40 & 16.4 & 25 & 2.16 & 0.066 & 0.022 & 0.015 & 0.003 & 0.42 & 0.41 & 0 & 0 & 0.137 & 1.00 & 0.99 & CA & 0.94 & 0.0029 & -- & -- \\ 
K02368.01 & 8.071 & N & 1.70 & 16.0 & 111 & 1.08 & 6.3e-05 & 1.4e-05 & 0 & 1.3e-06 & 0.0002 & 0.00011 & 2.3e-06 & 0 & 0.177 & 1.00 & 0.62 & CA & 0.0004 & 4.7e-05 & -- & 1210b \\ 
K02369.01 & 11.018 & N & 2.55 & 18.3 & 138 & 0.78 & 0 & 0 & 0 & 0 & 1.3e-05 & 0 & 8.2e-06 & 0 & 0.202 & 1.00 & 0.99 & CA & 2.2e-05 & 4.8e-06 & -- & 1211b \\ 
K02369.03 & 7.227 & N & 1.32 & 7.5 & 116 & 2.22 & 0 & 0 & 0 & 0 & 2.6e-05 & 0.00044 & 0.00015 & 1e-05 & 0.130 & 0.86 & 0.33 & CA & 0.00062 & 0.00012 & -- & -- \\ 
K02370.01 & 78.732 & N & 5.39 & 17.7 & 188 & 0.54 & 0.44 & 0.0027 & 0.074 & 0.0017 & 0.026 & 0.0012 & 0 & 0 & 0.051 & 1.00 & 1.00 & CA & 0.54 & 0.034 & -- & -- \\ 
K02371.01 & 12.941 & N & 2.14 & 19.4 & 85 & 1.74 & 8.1e-05 & 0 & 0 & 0 & 2.3e-05 & 0 & 1.1e-05 & 0 & 0.212 & 1.00 & 0.99 & CA & 0.00012 & 2.7e-05 & -- & 1212b \\ 
K02372.01 & 5.350 & N & 1.19 & 20.2 & 18 & 2.10 & 0 & 0 & 0 & 0 & 1.1e-06 & 0 & 2.8e-05 & 0 & 0.108 & 1.00 & 0.78 & CA & 3e-05 & 1.6e-06 & -- & 1213b \\ 
K02373.01 & 147.281 & N & 2.19 & 13.5 & 176 & 2.88 & 0.00042 & 3.4e-05 & 3.5e-06 & 3.3e-06 & 0.0017 & 0.004 & 0.00019 & 0 & 0.216 & 0.97 & 1.00 & CA & 0.0064 & 0.00052 & -- & -- \\ 
K02374.01 & 5.262 & N & 1.30 & 18.5 & 56 & 0.75 & 0 & 0 & 0 & 0 & 0.00012 & 0.00016 & 1.1e-05 & 1.3e-06 & 0.132 & 1.00 & 0.88 & PL & 0.00029 & 5.8e-05 & -- & -- \\ 
K02374.02 & 12.163 & N & 1.48 & 14.8 & 138 & 2.52 & 0.0012 & 0.004 & 6.2e-06 & 0.00058 & 0.3 & 0.18 & 0 & 0 & 0.153 & 1.00 & 0.56 & PL & 0.49 & 0.0055 & -- & -- \\ 
K02375.01 & 40.879 & N & 2.00 & 18.4 & 109 & 2.49 & 0 & 0.00026 & 0 & 6.9e-06 & 0.4 & 0.6 & 0 & 0 & 0.209 & 0.01 & 0.06 & FP & 0.99 & 0.0024 & -- & -- \\ 
K02376.01 & 18.826 & N & 2.36 & 17.1 & 229 & 1.23 & 0.0021 & 4.1e-06 & 9.7e-06 & 0 & 8.3e-05 & 4.3e-06 & 1.1e-05 & 0 & 0.214 & 1.00 & 0.90 & CA & 0.0023 & 0.00063 & -- & 1214b \\ 
K02377.01 & 13.903 & N & 1.50 & 15.0 & 131 & 0.90 & 2e-05 & 2.2e-06 & 0 & 1.5e-06 & 0.0025 & 0.00065 & 2e-06 & 0 & 0.151 & 0.07 & 0.88 & CA & 0.0031 & 0.00012 & -- & -- \\ 
K02378.01 & 4.767 & N & 1.22 & 12.8 & 65 & 1.23 & 0 & 6.4e-05 & 0 & 2.3e-05 & 0.0024 & 0.00051 & 3.6e-06 & 0 & 0.121 & 1.00 & 0.64 & CA & 0.0031 & 0.00016 & -- & 1215b \\ 
K02379.01 & 40.009 & N & 80.54 & 17.8 & 553 & 1.29 & 0.34 & 0.061 & 0.026 & 0.0089 & 0.37 & 0.2 & 0 & 0 & 0.001 & 1.00 & 0.98 & FP & 1 & 0 & -- & -- \\ 
K02380.01 & 6.357 & N & 1.93 & 15.6 & 102 & 0.78 & 0.033 & 0.014 & 0.0027 & 0.0019 & 0.014 & 0.014 & 1.4e-06 & 0 & 0.219 & 1.00 & 0.97 & CA & 0.081 & 0.0057 & -- & -- 
\enddata
\tablecomments{A portion of this table is shown for form and content.  
                The full table will be available online.}
\tablenotetext{a}{Maximum secondary eclipse depth allowed.}
\tablenotetext{b}{``Exclusion radius'' inside of which false positive scenarios are allowed.}
\tablenotetext{c}{Probabilities for different astrophysical false positive scenarios: 
                unblended eclipsing binary (EB), hierarchical eclipsing binary (HEB),
                and background/foreground eclipsing binary (BEB). "2" indicates double-period scenario.}
\tablenotetext{d}{Artificial models to identify signals that are poorly 
                described by any of the astrophysical scenarios.}
\tablenotetext{e}{Assumed ``specific planet occurrence rate'' for this planet.}
\tablenotetext{f}{Probability of signal to be on target star, according to Bryson et al.~(2015, in prep).}
\tablenotetext{g}{Positional probability score, from Bryson et al. (2015).}
\tablenotetext{h}{Exoplanet Archive disposition: false positive (FP), candidate (CA), or confirmed (PL).}
\tablenotetext{i}{False positive probability; mean of 10 bootstrap recalculations.}
\tablenotetext{j}{False positive probability uncertainty; standard deviation of 10 bootstrap recalculations.}
\tablenotetext{k}{Reason for failure: (1) No MCMC modeling available from \citet{Rowe:2015};
    (2)  Unphysical MCMC fit from \citet{Rowe:2015};
    (3)  No stellar parameters available from \citet{Huber:2014};
    (4)  No weak secondary data available;
    (5)  MCMC trapezoid fit did not converge;
    (6)  Period too short for implied star (orbit within star);
    (7)  Other unspecified \vespa\ error.}
\tablenotetext{l}{Kepler number assigned, if validated.}
\end{deluxetable*}

\end{turnpage}

\end{document}